\documentclass[12pt]{article}
\usepackage{amsfonts}
\usepackage{amssymb}
\usepackage{latexsym}
\usepackage{graphicx}
\usepackage[english]{babel}
\begin{document}
\input epsf

\renewcommand{\theequation}{\arabic{section}.\arabic{equation}}
\newcommand{\f}[2]{\frac{#1}{#2}}
\def\be{\begin{equation}}
\def\bea{\begin{eqnarray}}
\def\ee{\end{equation}}
\def\eea{\end{eqnarray}}
\def\c{\cosh\alpha}
\def\r{\rightarrow}
\def\pa{\partial}
\def\t{\tilde}
\def\nn{\nonumber\\ }
\def\g{\Gamma}
\def\h{{1\over 2}}
\def\ad{{\dot a}}
\def\b{\bigskip}
\def\m{\medskip}
\def\p{{\cal {P}}}
\def\q{{\cal {Q}}}
\def\qt{ {\tilde {\cal {Q}} }}
\def\s{{\cal {S}}}
\def\d{{d\over dt}}
\def\du{{d\over d\tau}}

\def\ph{\hat{\cal{P}}}
\def\qh{\hat{{\cal{Q}}}}
\def\qth{\hat{\tilde{\cal{Q}}}}
\def\sh{\hat{\cal{S}}}
\def\ih{\hat{I}}

\newcommand{\dt}[2]{\frac{d #1}{d #2}}

\begin{flushright}
\end{flushright}
\vspace{20mm}
\begin{center}
{\LARGE  Fractional Brane State in the Early Universe}\\
\vspace{18mm}
{\bf Borun D. Chowdhury\footnote{borundev@mps.ohio-state.edu} and Samir D. Mathur\footnote{mathur@mps.ohio-state.edu}}\\
\vspace{8mm}
Department of Physics,\\ The Ohio State University,\\ Columbus,
OH 43210, USA\\
\vspace{4mm}
\end{center}
\vspace{10mm}
\thispagestyle{empty}
\begin{abstract}

In the early Universe matter was crushed to high densities, in a manner similar to that encountered in gravitational collapse to black holes. String theory suggests that  the large entropy  of black holes can be understood in terms of fractional branes and antibranes. We assume  a similar physics for the matter in the early Universe, taking a toroidal compactification and letting branes wrap around the cycles of the torus. We find an equation of state $p_i=w_i\rho$, for which the dynamics can be solved analytically. For black holes, fractionation can lead to non-local quantum gravity effects across length scales of order the horizon radius; similar effects in the early Universe might change our understanding of Cosmology in basic ways. 

\end{abstract}
\newpage
\setcounter{page}{1}
\renewcommand{\theequation}{\arabic{section}.\arabic{equation}}
\section{Introduction}\label{introduction}
\setcounter{equation}{0}

At the beginning of the Universe we expect that energy density was high,  and it is likely that quantum gravity was important. Since string theory provides a consistent theory of quantum gravity, we must ask what kind of states are expected in string theory under these conditions. Several ideas have been considered, using either  string theory or string inspired constructions \cite{many,bran,greene1,greene2}. 

Another place where matter gets crushed to high densities is in the formation of a black hole.  In the classical picture of a black hole the curvature is low at the horizon, and large at the singularity. If we consider quantum mechanics on such a background geometry we run into the  the black hole information paradox \cite{hawking}, which implies that unitarity of quantum mechanics is lost.

String theory has made considerable progress in understanding black holes. We can understand the entropy of extremal and near extremal holes \cite{sen,sv,cm}, and obtain Hawking radiation as a unitary process where excited string states decay \cite{dmcompare}. This suggests that string theory will change our naive picture of the black hole geometry and allow information to leak out in the Hawking radiation.

Several computations have suggested a `fuzzball' picture of the black hole interior, where the quantum gravity effects are not confined to the vicinity of the singularity, but instead spread out all through the interior of the horizon. The key effect is `fractionation': when different kinds of branes are bound together they split up into fractional brane units \cite{dmfrac}. We can regard the large entropy  of the black hole as a consequence of fractionation: the entropy calculation just  counts these fractional brane units with their appropriate spins and fermion zero modes.  Fractionation is also responsible for the low energy of Hawking radiation quanta. More qualitatively, we can say that the fuzzball picture of the black hole interior is also a consequence of fractionation; fractional branes are low tension objects that can stretch to horizon scales instead of just planck distance from the singularity. The concrete computations leading to the fuzzball picture construct the microstates that account for the entropy. For 2-charge extremal holes we can understand all microstates, and  for 3 and 4 charge extremal cases subfamilies respecting one or more U(1) symmetries have been constructed \cite{microstates,micromore}. In each case the microstate is found to have been modified in the entire interior of the hole, and there is no horizon. If the fuzzball picture were true it would resolve the information paradox, since information can escape from the surface of the fuzzball, much like it leaves from the surface of a piece of burning coal. 

\begin{figure}[htbp] 
   \centering
   \includegraphics[width=2.5in]{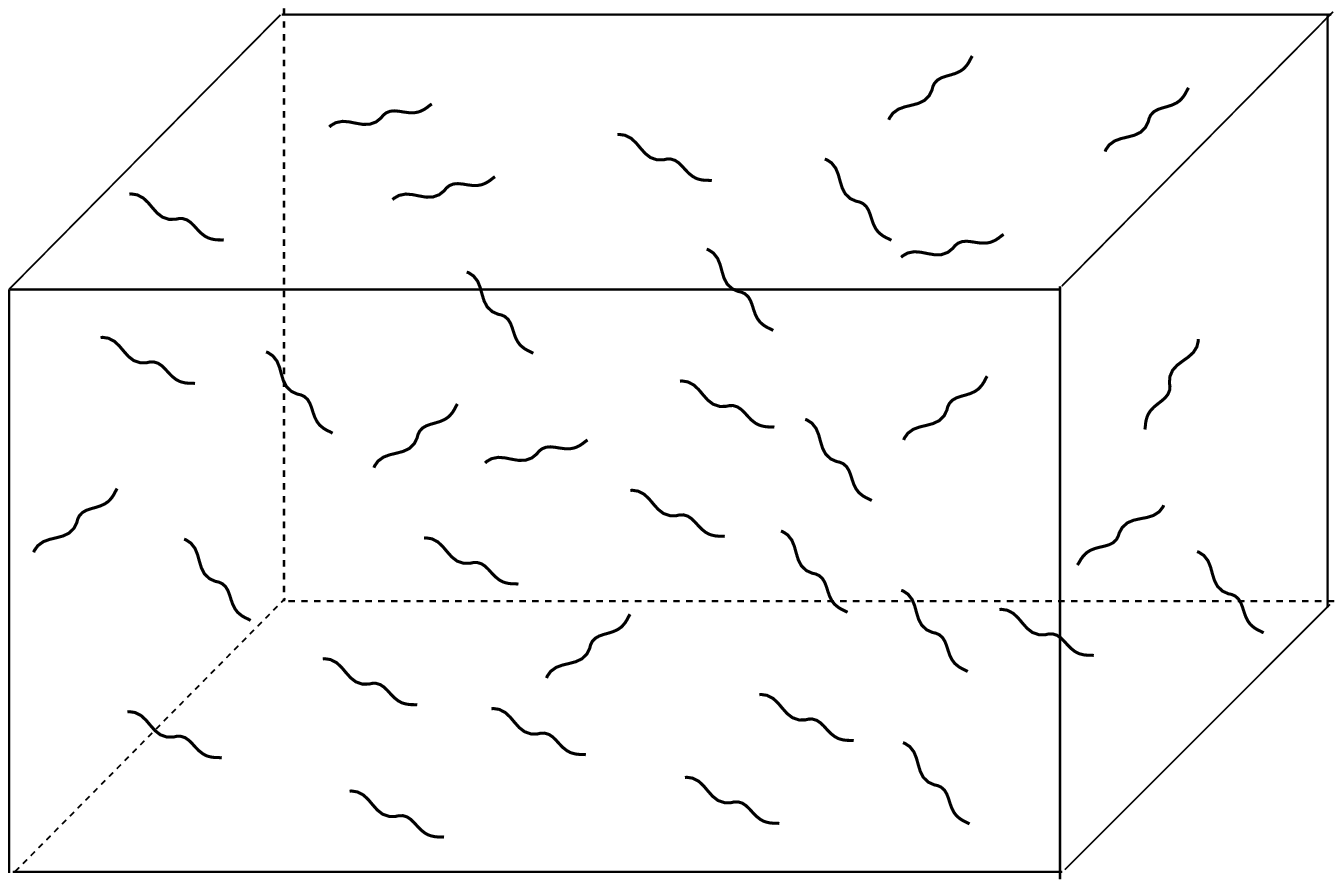} \hspace{1truecm}
    \includegraphics[width=2.5in]{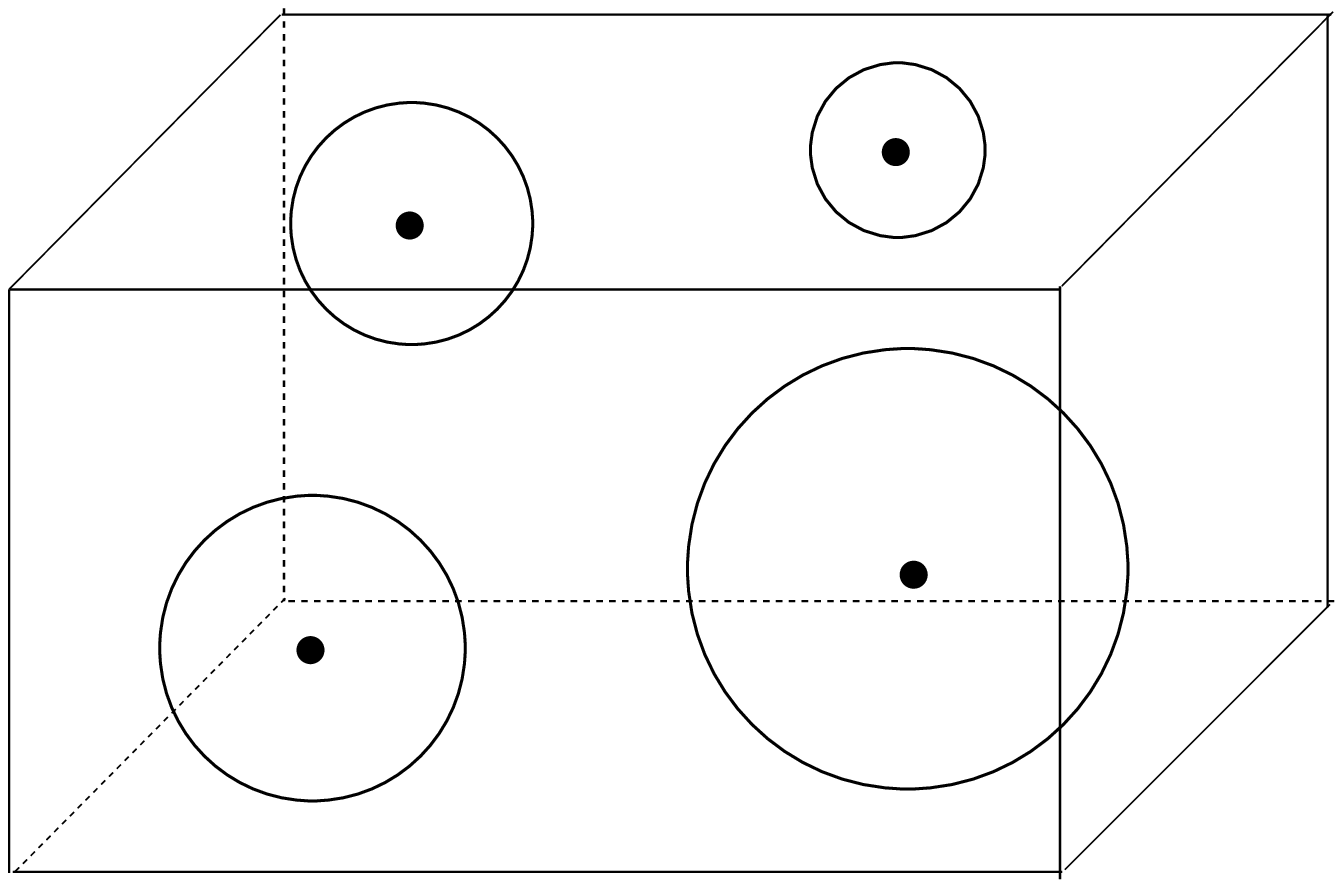} 
\vspace{.5truecm}
\hspace{4.5truecm} (a) \hspace{7truecm} (b)\\
\vspace{.5truecm}
 \includegraphics[width=2.5in]{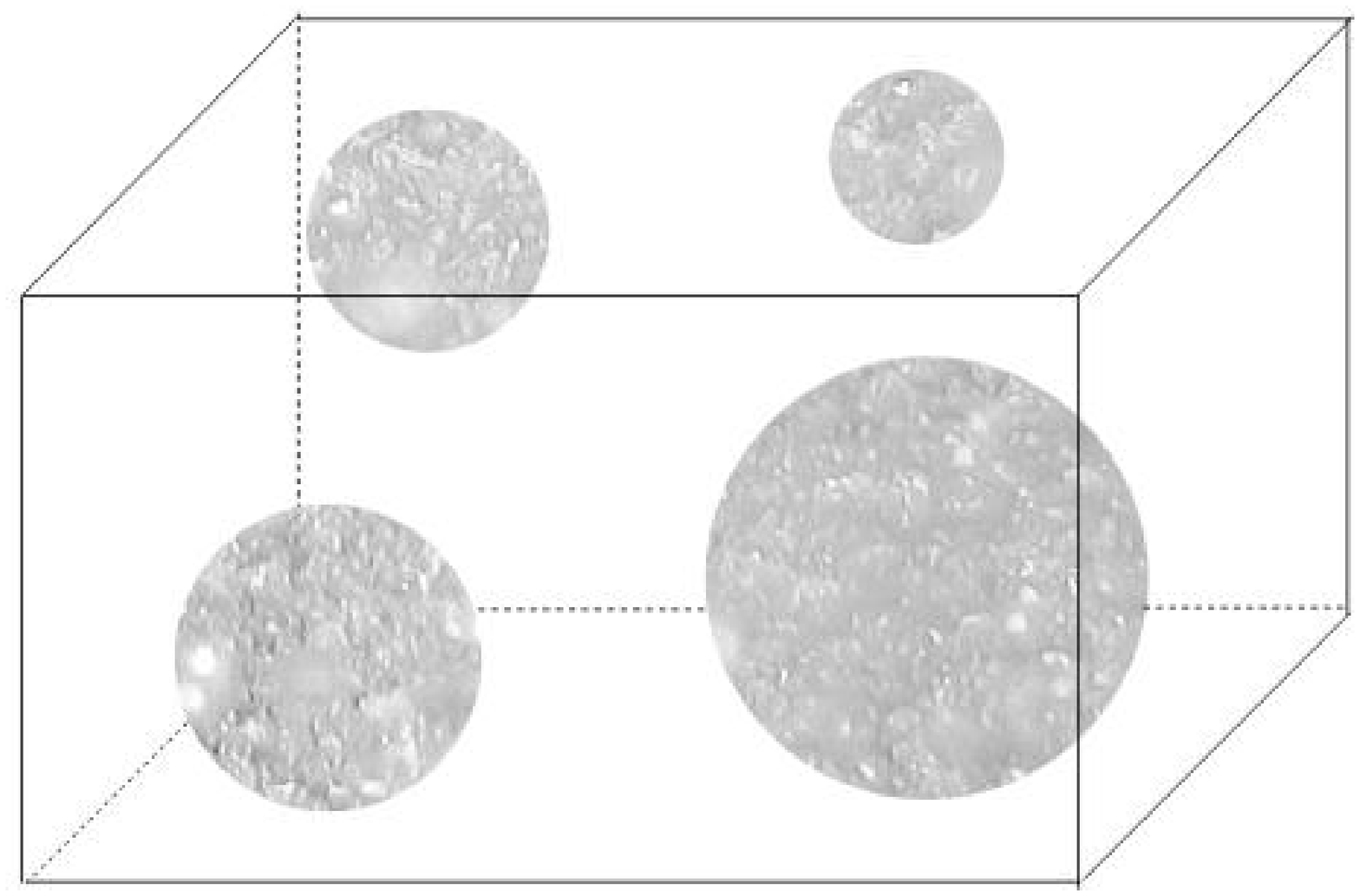}  \hspace{1truecm}
 \includegraphics[width=2.5in]{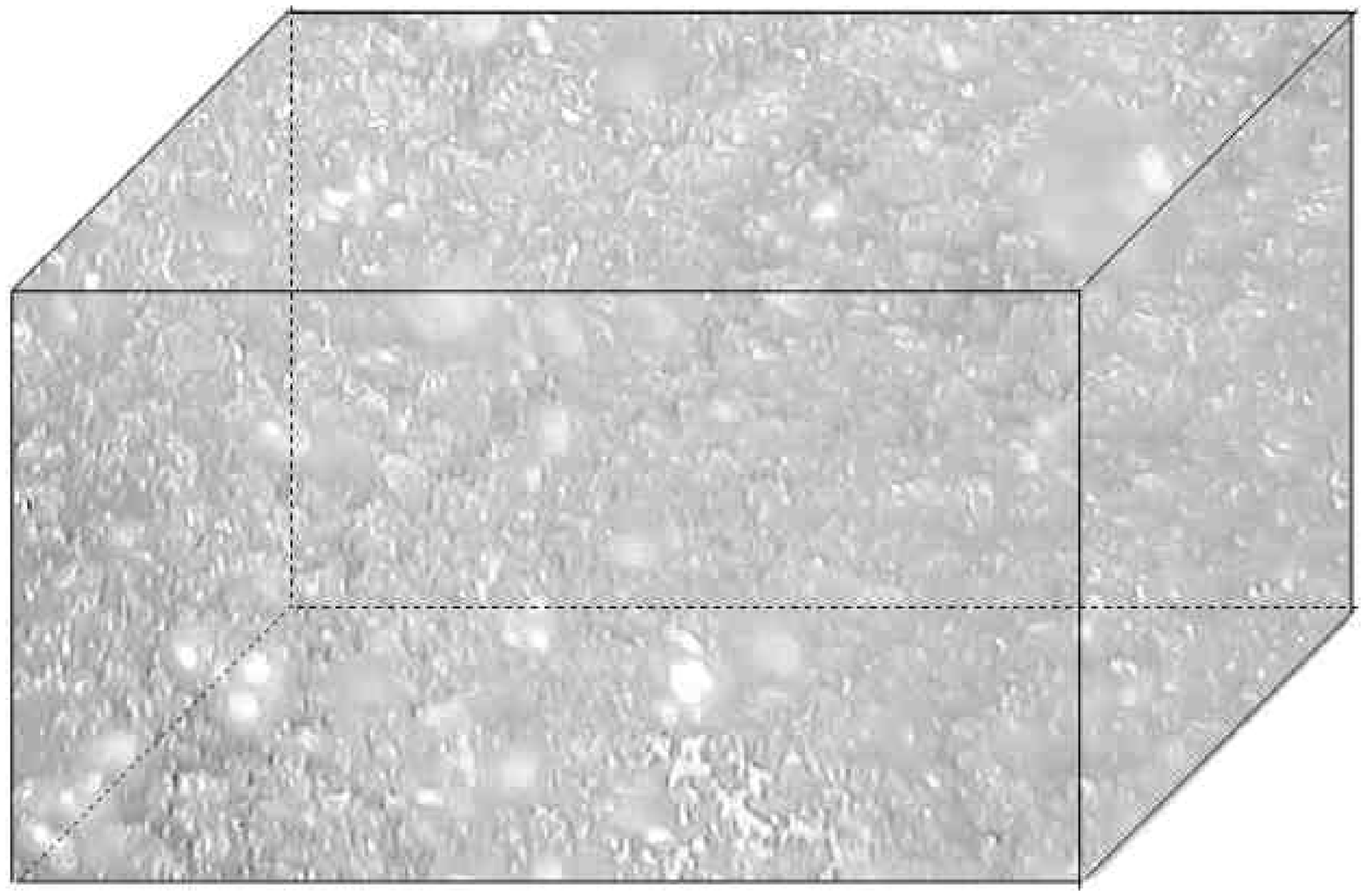} 
 \vspace{.5truecm}
\hspace{4.5truecm} (c) \hspace{7truecm} (d)
\caption{(a) Radiation filled Universe \quad (b) All matter in black holes \quad (c) Fuzzball picture suggests that interior of horizon is a very quantum domain \quad (d) Quantum fuzz filling the entire Universe.}
   \label{univ}
\end{figure}

In this paper we wish to ask the question: can we apply our understanding of black holes to say something about the Cosmological singularity? In Fig.\ref{univ}(a) we depict a traditional radiation filled Universe. We know that we can get a larger entropy for the same energy if we put the mass into black holes of sufficiently large radius; we depict this in Fig.\ref{univ}(b). But our Universe does not look like this at all; if black holes had formed at early times they would continue to exist till today (unless they were small enough to have Hawking evaporated by now).

The situation does not change in any material way if we replace the conventional picture of the black hole interior with a `fuzzball' (Fig.\ref{univ}(c)); this affects only the interior of the hole and not gross properties like the classical attraction between holes.

But if the maximal entropy state of a black hole is this quantum fuzz, then perhaps the maximal entropy state of the  Universe is given by such a  quantum  fuzz filling the entire Universe (Fig.\ref{univ}(d)). Using the microscopic expressions for black hole entropy we conjecture an equation of state for this fuzz, and find the evolution of the Universe with this equation of state.

\section{Fractional brane states and entropy}\label{entropy}
\setcounter{equation}{0}

In this section we will recall some results from the string description of black holes, which will motivate our ansatz of the fractional brane state and its entropy. A more detailed review of these results can be found in 
\cite{review}.

We will work with 10+1 dimensional M theory, using on occasion the language of 9+1 dimensional string theory when discussing branes. We will let the 10 space directions of M-theory be compactified to  $T^{10}$. We will denote the spacetime dimension as $D$.

Fig.\ref{univ}(a) depicts a Universe filled with radiation. M-theory has massless quanta, so we can certainly achieve such a state.  Let us fix the lengths of the sides of the torus, and explore the entropy as a function of the total energy. If the spacetime dimension is $D$ then
\be
S~\sim~ E^{D-1\over D}
\label{one}
\ee
Thus if the Universe was filled with massless radiation we would get $S\sim E^{10\over 11}$ for the 11 dimensions of M-theory and $S\sim E^{9\over 10}$ for the 10 dimensions of  string theory; in the latter case  $x^{11}$ has been compactified to a  small length so that quanta along $x^{11}$ are not excited.

\begin{figure}[htbp] 
   \centering
   \includegraphics[width=2.5in]{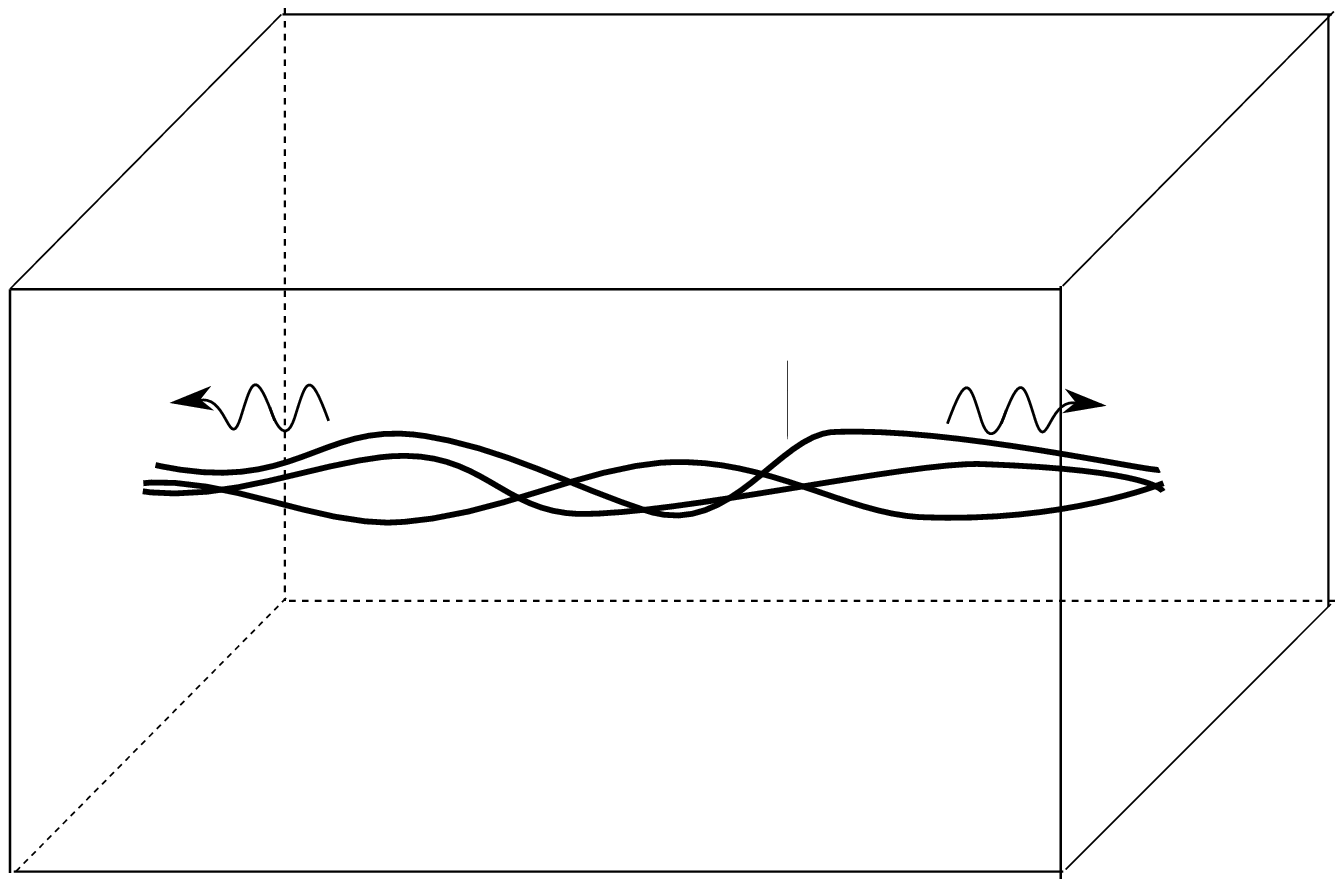} \hspace{1truecm}
   \includegraphics[width=2.5in]{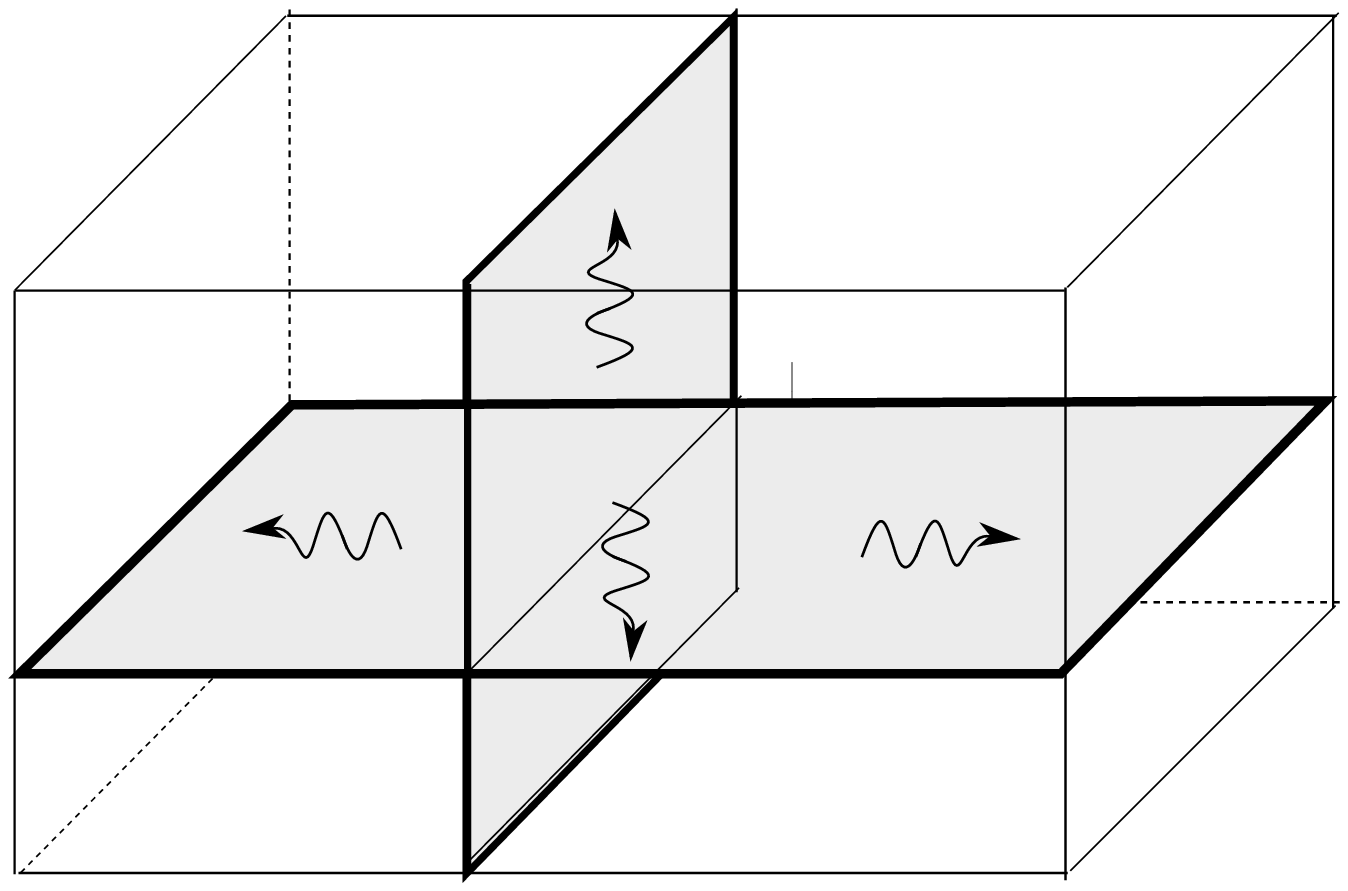} 
\vspace{.5truecm}
\hspace{4.5truecm} (a) \hspace{7truecm} (b)
\caption{(a) A string can wind several times around a compact cycle and carry vibrations \quad (b) In the `brane gas' model branes can wrap cycles and carry vibrations.}
   \label{strings}
\end{figure}

\subsection{Two charges}

Since we have extended objects in our theory, we can wrap them around the cycles of the torus. 
Consider string theory and let a string be wrapped $n_1$ times around a cycle of the torus; let the length of this cycle be $L$.  We can add excitations to this string, which split up into left movers and right movers. First let the string be a heavy `background' object, with the excitations as small vibrations. The excitations form a massless gas in $1+1$ dimensions. The total length of the string is $L_T=n_1 L$. The energy and momentum carried by the left movers is of the form
\be
E_L=|P_L|={\pi n_p\over L}={2\pi n_1n_p\over L_T}
\ee
The entropy of the left movers is
\be
S_L=2\sqrt{2}\pi\sqrt{n_1 n_p}
\ee
where the dependence $S\sim \sqrt{n_1n_p}$ comes from the way the momentum can be distributed among different harmonics on the string, and the coefficient arises from the fact that there are 8 transverse vibration modes of the string and 8 fermionic superpartners of these modes \cite{sen}. 
Note that the $n_p$ units of $P_L$ broke up into $n_1n_p$ `fractional' units of momentum because the string itself was a bound state of $n_1$ singly wrapped strings; this is a simple example of the fractionation mentioned above \cite{dmfrac}. 

Adding in the right movers we have
\be
S=2\sqrt{2}\pi\sqrt{n_1}~(\sqrt{n_p}+\sqrt{\bar n_p})
\label{two}
\ee
Of course we should not really regard the vibrations of the string as small oscillations in general, and to carry out the full computation we note that the total energy $E$ of a string state is given by
\be
E^2=(\hat n_1 L T-{2\pi \hat n_p\over L})^2+8\pi T N_L = (\hat n_1 L T+{2\pi \hat n_p\over L})^2+8\pi T N_R
\ee
where $T$ is the tension of the string, $\hat n_1=n_1-\bar n_1, \hat n_p=n_p-\bar n_p$ give the net winding and net momentum carried by the string,  and $N_L, N_R$ are the left and right excitation  levels. The entropy is 
\be
S=2\sqrt{2}\pi (\sqrt{N_L}+\sqrt{N_R})
\ee
For vanishing net winding and momentum $\hat n_1=0, ~\hat n_p=0$ we get
\be
S=2\sqrt{\pi} ~{E\over \sqrt{ T}}
\label{three}
\ee
This is a faster growth of S than (\ref{one}), and leads to the well known Hagedorn transition. 

We can understand the above dependence $S\sim E$ also in the more elementary computation 
(\ref{two}). The Universe will have no net string winding, so we will have winding as well as anti-winding modes. On the winding mode we have left movers (momentum) and right movers (anti-momentum), and similarly for the anti-winding modes. We find that the entropy is optimized if we put as much energy into string winding ($(n_1+\bar n_1) L T= 2 n_1 L T= {E\over 2}$) as in the momentum excitations. This gives 
\be
n_1=\bar n_1\sim E, ~~~n_p=\bar n_p\sim E, ~~~~S\sim \sqrt{n_1n_p}\sim E
\label{four}
\ee
in agreement with (\ref{three}).

The purpose of carrying out the estimate in the crude form (\ref{four}) is that we wish to talk about fractional branes and antibranes. In the count (\ref{three}) we had a closed loop of an excited string, but we see that we can regard the two sides of this loop as string `winding' and `antiwinding', and the excitations as `momentum' and `anti-momentum'. We have understood the state in terms of  two kinds of charges (and their anticharges):  windings of the elementary string (NS1) and momentum (P). We will call such states  `2-charge' states, and have found that the entropy of 2-charge states grows as $S\sim E$.

\begin{figure}[htbp] 
   \centering
   \includegraphics[width=3in]{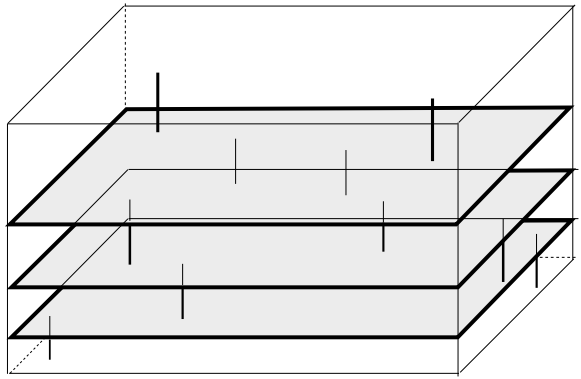} 
   \caption{Different kinds of branes  `fractionate' each other, giving a large entropy.}
   \label{fractionated}
\end{figure}

If the string has only left excitations but no right excitations then we get an extremal  NS1-P state.   The entropy is given by setting $\bar n_p=0$ in (\ref{two}), so we have $S=2\sqrt{2}\pi\sqrt{n_1n_p}$. We can use dualities to map this system to other forms. For example we can get D0-D4 -- a bound states of $n_0=n_1$ D0 branes and $n_4=n_p$  D4 branes. A further T-duality along a direction in the D4 gives $D1-D3$, where the $n_1$ D1 branes are perpendicular to the $n_3$ D3 branes.  We depict this in Fig.\ref{fractionated}. Note that each D1 brane gets  `broken up' into $n_3$ pieces. Thus there are $n_1n_3$ fractional D1 branes, and their different positions give $\sim n_1n_3$ moduli in a classical description of the branes. Quantizing the wavefunctions on this moduli space will again give the 2-charge entropy $S=2\sqrt{2}\pi\sqrt{n_1n_3}$.

\subsection{Three charges}

Can we get an entropy that grows with energy faster than $S\sim E$? Let us recall the microscopic description of 3-charge black holes. 

Consider type IIB string theory, and let there be 5 compact directions, which we write as $T^4\times S^1$:
\be
M_{9,1}~\rightarrow ~ M_{4,1}\times T^4\times S^1
\ee
We will wrap branes on the compact directions, and obtain an object that is a black hole in the  in 4+1 nonconmpact directions. The black hole in \cite{sv} was made with charges D1-D5-P, but since we started with the elementary string above we dualize this to get NS1-NS5-P. The NS1 branes are wrapped on $S^1$, the NS5 branes wrap $T^4\times S^1$, and the momentum P runs along $S^1$. The entropy is \cite{sv}
\be
S=2\pi\sqrt{n_1n_5n_p}
\ee
Let the mass of a brane of type $i$ be $m_i$. Then the energy of the extremal system is just given by adding the masses of the branes
\be
E=n_1m_1+n_5m_5+n_pm_p
\ee
Since the energy of the system is linear in the numbers of branes $n_i$, we find
\be
S\sim E^{3\over 2}
\label{ten}
\ee
Thus the 3-charge entropy grows faster with energy than the 2-charge entropy (\ref{three}). 

Suppose  $n_1, n_5\gg n_p$. Let us add a small amount of extra energy without changing the charges, so that the system is no longer extremal. One finds that the Bekenstein entropy of the near-extremal hole $S(E, Q_i)$ can be reproduced by assuming that we have both momentum and anti-momentum excitations, and that these do not interact \cite{cm}:
\be
S=2\pi\sqrt{n_1n_5}(\sqrt{n_p}+\sqrt{\bar n_p})
\label{threenex}
\ee
\be
E=n_5m_5+n_1 m_1+(n_p+\bar n_p)m_p
\ee
If only one charge is large, $n_5\gg n_1, n_p$ and the system is slightly off extremality, we find that we can again reproduce the Bekenstein entropy exactly \cite{malda5} by writing
\be
S=2\pi\sqrt{n_5}(\sqrt{n_1}+\sqrt{\bar n_1})(\sqrt{n_p}+\sqrt{\bar n_p})
\label{five}
\ee
\be
E=n_5m_5+(n_1+\bar n_1) m_1+(n_p+\bar n_p)m_p
\label{six}
\ee
The expression (\ref{five}) needs some explanation. The number $n_5$ is determined by the given NS5 charge  for the system, but there are four other numbers: $n_1, \bar n_1, n_p, \bar n_p$. The net NS1 and P charges ($\hat n_1, \hat n_p$) give
\be
n_1-\bar n_1=\hat n_1, ~~~~n_p-\bar n_p=\hat n_p
\ee
and a third relation comes from the energy (\ref{six}). This leaves one free parameter, and we should extermize $S$ over this parameter to obtain the entropy. The result then tells us how the energy wants to partition itself into fractional excitations, and is found to exactly reproduce the Bekenstein entropy of the system. 

We can make a natural extension of the above formulae for entropy to the case where all charges are comparable, and the system is not close to extremal. This case will therefore include the Schwarzschild hole.  We write \cite{hms}
\be
S=2\pi(\sqrt{n_5}+\sqrt{\bar n_5})(\sqrt{n_1}+\sqrt{n_1})(\sqrt{n_p}+\sqrt{\bar n_p})
\label{el}
\ee
\be
E=(n_5+\bar n_5)m_5+(n_1+\bar n_1) m_1+(n_p+\bar n_p)m_p
\label{seven}
\ee
Again we have the conditions
\be
n_5-\bar n_5=\hat n_5, ~~~~n_1-\bar n_1=\hat n_1, ~~~~n_p-\bar n_p=\hat n_p
\ee
with the energy given by (\ref{seven}) which assumes no interaction energy between the branes and antibranes. This time there are 6 parameters and 4 conditions, and we must again extremize $S$ over the remaining 2-parameter family to get the correct $S$. The resulting $S(E, Q_i)$ agrees exactly with the Bekenstein entropy of black holes in 4+1 dimensions, for all values of charges $Q_i$ and energy $E$. 

If we scale up all charges and the total energy the same way, we find that for all these entropies arising from using three kinds of charges we get $S\sim E^{3\over 2}$.

\subsection{4 charges}

A similar story is found if we compactify an additional direction, so that we have
\be
M_{9,1}~\rightarrow ~ M_{3,1}\times T^4\times S^1\times \tilde S^1
\ee
Now we have 6 directions on which we can wrap objects, and we get black holes in 3+1 noncompact dimensions. We can take the NS1-NS5-P charges that we had above and add a fourth charge: KK monopoles that have $\tilde S^1$ as their non-trivially fibred circle. We can again dualize these four charges to a variety of forms. A form that looks symmetric in the four charges consists of four D3 branes that wrap the 6 cycles of the compact $T^6$ as follows
\be
D3_{123}~~~~D3_{145}~~~~D3_{246}~~~~D3_{356}
\ee
Thus any pair of D3 branes shares one common direction. 
The entropy is again given by \cite{4charge}
\be
S=2\pi(\sqrt{n_1}+\sqrt{\bar n_1})(\sqrt{n_2}+\sqrt{\bar n_2})(\sqrt{n_3}+\sqrt{\bar n_3})(\sqrt{n_4}+\sqrt{\bar n_4})
\label{eight}
\ee
where we extremize over the $n_i, \bar n_1$ subject to
\be
n_i-\bar n_i=\hat n_i
\ee
and
\be
E=\sum_i (n_i+\bar n_i) m_i
\ee
This entropy agrees exactly with the Bekenstein entropy $S(E,Q_i)$ of a hole in 3+1 dimensions.

By changing the orientation of one of the D3 branes we get a nonsupersymmetric but still extremal system, and the entropy of this system was matched to the Bekenstein entropy recently in \cite{emho}. 

Note that the $n_i$ grow linearly with $E$, so the entropy (\ref{eight}) grows with energy as
\be
S\sim E^2
\label{nine}
\ee

\subsection{Proposal for entropy in the early Universe}

We have seen that if we take a gas of massless particles we get an entropy $S\sim E^{D-1\over D}$. We can consider excited string states which give $S\sim E$; we have seen that this system can be re-interpreted as a 2-charge system where the two charges fractionate each other and produce the entropy. With three charges we get $S\sim E^{3\over 2}$. With four charges we get $E\sim E^2$.

Now consider the Universe, where we have compactified all the spatial directions to a torus. The traditional big bang picture envisages a radiation filled Universe at early times. In \cite{bran} a string gas was considered and in \cite{greene1,greene2} a `brane gas' was taken. Such gases can give entropy $S\sim E$. But we have seen above that general fractional brane states can give a much higher $S$ at large $E$. We wish to adopt  an equation of state for   the early Universe that will reflect this
high entropy and thus correspond to the most generic state for given $E$.

We will assume that the entropy has the form
\be
S=A'\prod_{i=1}^N (\sqrt{n_i}+\sqrt{\bar n_i})
\label{entropyass}
\ee
Since we will take the net charge of each type to vanish,  we have $n_i=\bar n_i$ and we get
\be
S=2^N A' \prod_{i=1}^N \sqrt{n_i}\equiv  A\prod_{i=1}^N \sqrt{n_i}
\ee
The energy is
\be
E=\sum_i m_i (n_i+\bar n_i)=2\sum_i m_i n_i
\ee
Here $m_i$ is the mass of the brane of type $i$  
\be
m_i=T_p \prod_j L_j
\ee
where $T_p$ is the tension of a p-brane and the product runs over all the spatial directions of the brane.

We assume that the system is in thermal equilibrium, so we will maximize the entropy (\ref{entropyass}) for given $E$.\footnote{To see if the assumption of equilibrium is true, we will have to compute the rate of interactions between fractional branes. This interaction depends on the total number of branes in the bound state. We do not address these issues here, and hope to return to them elsewhere.} To find the state with maximal entropy at a given time $t$, the $L_i$ are held fixed (which fixes the $m_i$), and the total energy is held fixed at $E$. Taking into account this energy constraint we maximize
\be
\t S=S-\lambda (E_{branes}-E)= A \prod_{i=1}^N \sqrt{n_i}-\lambda( 2 \sum_i m_i n_i-E)
\ee
Extremizing over $n_i$ gives
\be
n_k=\bar n_k={E\over 2N m_k}
\ee
Note that the energy is equipartitioned among all types of branes, each type getting energy (there is no sum over $k$)
\be
E_k=n_km_k={E\over 2N}
\label{tw}
\ee

\subsection{Stress tensor}

We have seen that the entropy of black holes is reproduced by assuming that the energy gets partitioned optimally between different kinds branes and antibranes. In this computation the energy is taken to be just  additive; i.e. there was no energy of interaction. In \cite{hms} it was shown that with this same assumption of noninteraction between branes we can reproduce the {\it pressures} exerted by the black hole on the various compact cycles. Thus on the one hand we can take the black hole geometry and for   compact directions $y_i$  look at the asymptotic fall-off of $g_{y_iy_i}$; this is related to the pressure components $T^i_i$ of the stress tensor in a weak gravity situation.  On the other hand we can take the set of branes and antibranes that we obtained by extremizing an expression like (\ref{el}),(\ref{eight}),  compute the pressure each brane exerts by itself on the compact directions, and just add these pressures. One again finds exact agreement between the black hole result and the microscopic computation.\footnote{In \cite{hms} the variables compared between the two computations were certain  linear combinations of the pressures; for a direct computation of pressures from wrapped branes see for example \cite{cgm}. } We will thus also use a simple sum over the pressures of the branes describing our configuration. 

Let us first compute the stress tensor of a single $p$-brane. The action of the brane is
\be
S=-T_p~\int ~\sqrt{-g^{ind}} ~d^{p+1}\xi
\ee
where $g^{ind}_{ab}$ is the metric induced on the worldvolume. The stress tensor is given by
\be
T_{\mu\nu}=-{2\over \sqrt{-g}}{\delta S\over \delta g^{\mu\nu}}
\ee
Let the length of the direction $x^i$ be $L_i$. Let the brane be wrapped on directions $x^1\dots x^p$. The volume of the brane is  $V_p=\prod_{i=1}^p L_i$. The volume of the directions transverse to the brane is $V_{tr}=\prod_{i=p+1}^{D-1} L_i$. The total volume of the torus is $V=V_p V_{tr}$.
The stress tensor has only diagonal components. We find (there is no sum over $k$)
\bea
T^{(p) k}{}_k=-T_p ~\prod_{i=p+1}^{D-1}\hat \delta(x_i-\bar x_i), ~~~~~~~~~&k&=1, \dots, p\nn
T^{(p) k}{}_k=0, ~~~~~~~~~~~~~~~~~~~~~~~~~~~~~~~~~~&k&=p+1, \dots, (D-1)
\eea
where $\hat\delta$ is the covariant delta function ($\int \hat\delta(x) \sqrt{-g_{xx}}~ dx=1$), and $\bar x_i$ give the position of the $p$-brane in the transverse coordinates. 

Now suppose there are $n_p$ branes of this type, smeared uniformly on the transverse directions $x_i, i=p+1\dots (D-1)$. Then we get
\be
T^{(p) k}{}_k=-T_p {n_p\over V_{tr}}=-T_p {n_pV_p\over V}= - {E_p\over V}
\ee
where $E_p=n_p T_p V_p$ is the total energy carried by this type of brane. Using (\ref{tw})
we have
\be
T^{(p) k}{}_k =- {E\over 2NV}= - {\rho\over 2N}
\ee
where $\rho={E\over V}$ is the energy density. Including the contribution of the corresponding antibrane, we get from this type of brane the pressure
\be
p=-{1\over N}\rho
\ee
Now suppose there were $N_i$ types of branes wrapping the direction $x_i$. Then the pressure in the direction $x_i$ will be
\be
p_i=-{N_i\over N}\rho\equiv w_i\rho
\label{wncq}
\ee
where we have defined
\be
w_i\equiv -{N_i\over N}
\label{wnc}
\ee
 Momentum modes P contribute $-1$ to $N_i$ (they have a positive pressure while branes have a negative pressure). Note that the $N_i$ appearing in (\ref{wnc}) counts the {\it types} of branes wrapping different cycles, not the {\it number} of branes along those cycles. An example might make this clearer. Take 11-dimensional M theory; then there are 10 compact spatial directions. Consider the charges
 \be
 M5_{12345}, ~~~~M5_{12367}, ~~~~M5_{14567}, ~~~~P_1
 \ee
 where the subscripts indicate the directions along which the branes wrap ($P_1$ is momentum along the direction $x_1$ common to all M5 branes). There are 4 kinds of charges; thus $N=4$. Along $x_1$ we have the contribution $+1$ from each of the M5 branes and the contribution $-1$ from P; thus we have $N_1=3-1=2$, and $w_1=-{N_1\over N}=-{1\over 2}$. For $x_2$, we have a contribution $+1$ from the first and second types of M5 branes, so we again get $N_2=2$ and $w_2=-{1\over 2}$. A similar result holds for $x_3, \dots , x_7$. Along $x_8, \dots , x_{10}$ we find no charges, so $N_i=0$ and the corresponding $w_i$ vanish. Thus we get 
 \be
\{w_1, \dots , w_{10}\}\equiv  \vec w = \{ -\h, -\h, -\h,   -\h, -\h, -\h,   -\h, 0,0,0\}
 \ee
 
\subsection{Comparison between our approach and brane gas models}\label{branesec}

What new features can strings and branes bring to the early Universe?
In \cite{bran} the idea of `string gas' was examined; in \cite{greene1,greene2} the extension to `brane gases' was considered.
We will also be using branes, and some of our computations will resemble those in brane gas scenarios. But there is a significant difference in the basic idea between our approach and brane gases. Here we outline some points of this difference; this discussion should help us put forth our conjectured picture more clearly. 

\bigskip

(a) In a string gas we can wrap strings along arbitrary cycles of the torus. Similarly, in the brane gas of \cite{greene2} M2 branes were wrapped on all cycles $(ij)$ of the torus. But in our computation we need to take a set of branes that are mutually BPS; it is only such sets
that manifest fractionation and the consequent large entropy (\ref{entropyass}). M2 branes are mutually BPS if they share no common directions, M5 branes need to share three common directions,
and an M2 -  M5 combination must share one common direction.

\bigskip

(b) In the brane gas model of \cite{greene2} the entropy comes from vibrations living on the surface of the M2 branes. This entropy is proportional to the area of the brane, and at a suitably high temperature  the energy cost of each unit area of the brane is balanced by the entropy carried by that area. This gives a Hagedorn phase with $S\sim E$.

In our case the energy $E$ goes to creating certain sets of branes and antibranes that fractionate each other, and thereby create the entropy
(\ref{entropyass}). Even if we have just three kinds of charges, this entropy grows as $S\sim E^{3\over 2}$, much faster than the Hagedorn entropy. Thus at high energy densities we expect to get the fractional brane state rather than a Hagedorn state. 

\bigskip

(c) In the brane gas model when two branes intersect they tend to annihilate. Thus sets of branes that  do not generically intersect  are expected to last for longer times, and govern the long time dynamics of the system.

The situation is quite the opposite in our case. Consider the three charges NS1-NS5-P which we considered for the 3 charge black hole. The NS1 is bound to the NS5, and thus `lives in the plane of the NS5'. The P charges are carried by excitations of  the NS1-NS5   bound state along the direction common to the NS1 and NS5. For fractionation to occur, all charges must `see'  each other. 

One may then wonder why the branes and antibranes do not immediately annihilate to radiation. But we have already seen that if $E$ is high then there is more entropy in the fractional brane state than in radiation, so there should {\it not} be an annihilation to radiation. Let us discuss the annihilation of brane anti-brane pairs in more detail.

If we place a brane and an antibrane together, we get a system that can be described by a tachyon sitting at the top  of its potential hill \cite{sent}. Classically the tachyon can sit at the top of the hill indefinitely, while quantum mechanically it will fall to its ground state. If we take a large number $N$ of branes, and one antibrane, then we can describe the branes by their gravity dual. We then find that the antibrane just falls down the throat created by the branes, and no radiation emerges for long times \cite{tachyon}; thus there is no quick annihilation between the branes and the antibrane.

In fact  the annihilation process for fractional branes has been well studied in the black hole context, where one finds that  decay of brane-antibrane pairs gives Hawking radiation. For the three charge system described by (\ref{threenex}) we can compute the rate of annihilation of $P\bar P$ pairs to radiation, and find that the rate matches Hawking emission exactly in spin dependence and grey body factors \cite{dmcompare}. Similar agreement is found for the 4-charge analogue of (\ref{threenex}) \cite{klebanov} and  for the system described by (\ref{five}) \cite{mk}. We can even get the exact emission rate for the general hole described by (\ref{el}),(\ref{eight}), if we  boost the neutral hole to add charges; this maps the neutral hole    to the near-extremal system used in the above mentioned results \cite{ramadevi}. While boosting in a compact direction is not an exact symmetry of string theory, it may be a good approximation for large charges, and is similar to the idea used in Matrix theory.

All these computations suggest the following picture for black hole microstates. The state has a large number of fractional branes and antibranes, and the potential describing this system has a large number of saddle points which give metastable states.The system slowly drops from one metastable state to a lower energy one, giving 
Hawking radiation, which is a process suppressed by powers of $\hbar$ for classical sized black holes. We thus expect  that our system  on the torus $T^{10}$ will be composed of fractional branes and antibranes with the high entropy (\ref{entropyass}), and branes and antibranes will not annihilate.\footnote{Brane-antibrane models for black holes were also considered in \cite{fractional}.}

\bigskip

In summary,  let us take an analogy from nuclear physics. At low energy we see hadrons, but at high density and pressure we get a quark-gluon plasma, where deconfinement has liberated the elementary degrees of freedom to generate the highest possible entropy. At very high energies these elementary constituents are essentially noninteracting quanta. In our case we have a high energy density in the early Universe, and black hole physics suggests that the most entropically favored configuration is one of fractional branes. Black hole computations also suggest that these fractional brane quanta are free to leading order, and that we should find the total energy and pressure by adding the contributions from each brane in the state.

\section{Einstein's equations}
\setcounter{equation}{0}

We take the metric to have the form
\be
ds^2=-dt^2+\sum_{i=1}^{D-1}a^{2}_{i}(t)dx_{i}^{2}
\label{metric}
\ee
The coordinates $x_i$ are compactified with period unity ($0\le x_i<1$). The nonvanishing components of the connection are
\be
\g^t_{ii}=a_i\ad_i, ~~~\g^i_{ti}={\ad_i\over a_i}
\ee
The relevant components of the Einstein tensor are
\be
G^t{}_t=-\h (\sum_i{\ad_i\over a_i})^2+\h \sum_i {\ad_i^2\over a_i^2}
\ee
\bea
G^k{}_k&=&{\ddot a_k\over a_k} +{\ad_k\over a_k}(\sum_{i}{\ad_i \over a_i})-{\ad_k^2\over a_k^2}-\h [2\sum_i {\ddot a_i\over a_i}+(\sum_i{\ad_i\over a_i})^2-\sum_i {\ad_i^2\over a_i^2}]\nn
&=&{\ddot a_k\over a_k} +{\ad_k\over a_k}(\sum_{i}{\ad_i \over a_i})-{\ad_k^2\over a_k^2}- \sum_i {\ddot a_i\over a_i}+G^t{}_t
\label{gkkl}
\eea
(There is no sum over $k$ in (\ref{gkkl}).)
The Einstein equations are $G^\mu{}_\nu=8\pi G T^\mu{}_\nu$. The nonvanishing components of the stress tensor are
\be
T^t{}_t=-\rho, ~~~~~T^k{}_k=p_k=w_k\rho
\ee
so we get the field equations
\be
-\h (\sum_i{\ad_i\over a_i})^2+\h \sum_i {\ad_i^2\over a_i^2}=-8\pi G\rho
\label{gtt}
\ee
\be
{\ddot a_k\over a_k} +{\ad_k\over a_k}(\sum_{i}{\ad_i \over a_i})-{\ad_k^2\over a_k^2}- \sum_i {\ddot a_i\over a_i}=8\pi G (1+w_k)\rho
\label{gkkpre}
\ee
Substituting (\ref{gtt}) in (\ref{gkkpre}) we get
\be
{\ddot a_k\over a_k} +{\ad_k\over a_k}(\sum_{i}{\ad_i \over a_i})-{\ad_k^2\over a_k^2}- \sum_i {\ddot a_i\over a_i}=(1+w_k)~[\h (\sum_i{\ad_i\over a_i})^2-\h \sum_i {\ad_i^2\over a_i^2}]
\label{gkk}
\ee

\section{A Kasner type  power law solution}\label{kasner}
\setcounter{equation}{0}

For the empty Universe with toroidal compactification we have the Kasner solutions \cite{kasner}, where the radii grow as powers of $t$. A power law solution was also found for the case of isotropically wrapped branes in \cite{greene1}. We will see that with the equation of state that we have chosen we can get a power law solution for any choice of the $w_i$ which characterize the brane wrappings.

Thus write
\be
a_i=\bar a_i ~t^{\beta_i}
\ee
Thus
\be
{\ad_i\over a_i}={\beta_i\over t}, ~~~~~{\ddot a_i\over a_i}={\beta_i(\beta_i-1)\over t^2}
\ee
Substituting in (\ref{gkk}) gives
\be
\beta_k={\h (\sum_{i}\beta_i^2)(1-w_k)+\h(\sum_{i }\beta_i)^2(1+w_k) -(\sum_i\beta_i)\over  [(\sum_{i}\beta_i)-1]}
\ee
We write
\be
\sum_i \beta_i=A, ~~~~ \sum_i\beta_i^2=B
\label{sixt}
\ee
Then we have
\be
\beta_k=[{\h B+\h A^2-A\over (A-1)}]-w_k~ [{\h B-\h A^2\over (A-1)}]
\label{thir}
\ee
Let us define
\be
W\equiv \sum_i w_i, ~~~~~~U\equiv\sum_i w_i^2
\ee
We can get two consistency conditions from (\ref{thir}). First we sum over $k$ in (\ref{thir}), getting
\be
\sum_k\beta_k=A=(D-1)[{\h B+\h A^2-A\over (A-1)}]-W~[{\h B-\h A^2\over (A-1)}]
\label{fourt}
\ee
Next we square the $\beta_k$ and then add:
\be
\sum_k\beta_k^2=B=(D-1) [{\h B+\h A^2-A\over (A-1)}]^2+U[{\h B-\h A^2\over (A-1)}]^2-2W~[{\h B+\h A^2-A\over (A-1)}]~[{\h B-\h A^2\over (A-1)}]
\label{fift}
\ee
One solution to these equations is $A=1, B=1$, which gives the well known vacuum Kasner solutions \cite{kasner}. To find other solutions, note that 
eq.(\ref{fourt}) is linear in $B$, and gives
\be
B=A~{2(D-2)+A(3-W-D)\over D-1-W}
\ee
Substituting in (\ref{fift}) we get a quadratic equation for $A$. Solving this, we get two additional solutions, one of which is $A=0$. Collecting all these solutions we have the following cases:

\b

(i)
\be
A=0, ~~~~B=0
\ee
This gives $\beta_i=0$ for all $i$, and thus corresponds to empty Minkowski space.

\medskip

(ii)
\be
A=1, ~~~~B=1
\ee
These are the known  vacuum Kasner solutions. Thus there will be no matter, and the different expansions of the different directions give a self-consistent solution of the Einstein equations.
All $\beta_i$ satisfying (\ref{sixt})  with $A=B=1$ give allowed solutions.

\medskip

(iii)
\bea
A&=&{2(D-1-W)\over (D-1)+(D-2)U-W^2}\nn
B&=&4~{(D-1)+(D-2)^2U-2W-(D-3)W^2\over [(D-1)+(D-2)U-W^2]^2}
\eea
This gives a  solution with a nontrivial stress tensor contributed by branes. From (\ref{thir}) we find
\bea
\beta_k&=&[{\h B+\h A^2-A\over (A-1)}]-w_k~ [{\h B-\h A^2\over (A-1)}]\nn
&=&[{2(W-1)\over W^2-D(U+1)+2U+1}]-w_k~[{2(D-2)\over W^2-D(U+1)+2U+1}]
\label{betak}
\eea

\section{The equations in the general case}
\setcounter{equation}{0}

Let us write
\be
\gamma_i \equiv \f{\dot{a}_i}{a_i}
\label{gdef}
\ee
Thus
\be
\f{\ddot{a}_i}{a_i}=\dot{\gamma}_i + \gamma_i^2
\ee
Equation (\ref{gkk}) gives
\be
\dot{\gamma}_k +  \gamma_k (\sum_i \gamma_i) -\sum_i (\dot{\gamma}_i + \gamma_i^2) =\h \left[ (\sum_i \gamma_i)^2 - \sum_i \gamma_i^2 \right] (1+ w_k)
\label{eightt}
\ee
Let us define
\bea
\p&\equiv& \sum_i \gamma_i\nn
\q&\equiv& \sum_i \gamma_i^2\nn
\s&\equiv & \sum_i w_i \gamma_i
\label{tfour}
\eea
Then (\ref{eightt}) is
\be
\d{\gamma}_k + \gamma_k \p-\d{\p}-\q=\h(\p^2-\q)(1+w_k) 
\label{master}
\ee
Summing (\ref{master}) over $k$ gives
\be
-(D-2) \d\p+\p^2 -(D-1)\q =\h(\p^2-\q)(D-1+W)
\label{tone}
\ee
Multiplying (\ref{master}) by $\gamma_k$ and then summing over $k$ gives
\be
\h \d\q-\p\d\p =\h(\p^2-\q)(\p+\s)
\label{ttwo}
\ee
Multiplying (\ref{master}) by $w_k$ and then summing over $k$ gives
\be
\d \s+\p\s-W\d\p-W\q=\h(\p^2-\q)(W+U)
\label{tthree}
\ee
Interestingly, we find that even though there are $D-1$ variables $\gamma_i$, the three moments (\ref{tfour}) form a closed system of three first order equations. We can write (\ref{tone})-(\ref{tthree}) in a more convenient form by defining
\be
\qt=\q-\p^2
\ee
Then our three equations become
\bea
\dot \p~+~\p^2&=&- K_1 \qt \label{tsix}\\
\dot \qt~+~\p\qt&=&-\s \qt \label{tseven}\\
\dot \s~+~\p\s&=&K_2 \qt \label{teight}
\eea
where
\bea
K_1&=&{(D-1-W)\over 2(D-2)}\label{k1}\\
K_2&=&-\h[{1-W\over D-2}~W+U]
\label{k2}
\eea
If $\p, \q, \s$ are known then we get the $\gamma_i$ from (\ref{master})
\be
\dot\gamma_k+\gamma_k \p=-\h\qt [{1-W\over D-2}+w_k]
\label{qsix}
\ee
The $a_i$ are then determined by (\ref{gdef}).

\subsection{A more convenient form of the equations}

The left hand sides of (\ref{tsix})-(\ref{teight}) have a similar form, which suggests that we define an integrating factor. Consider eq.(\ref{teight}). We can write it as
\be
{d\over dt} (e^{\int_{t_0}^t \p dt} \s )=K_2 e^{\int_{t_0}^t \p dt}\qt
\label{tnine}
\ee
where $t_0$ is an arbitrary constant that we will take as the initial time where we specify data. 
For any quantity $F$ we write
\be
\hat F~\equiv~e^{\int_{t_0}^t \p dt}~F
\label{defhat}
\ee
Then  (\ref{tnine}) becomes
\be
\d \sh=K_2 \qth
\ee
Similarly, eq.(\ref{tsix}) becomes
\be
\d \ph=-K_1 \qth
\ee
Eq.(\ref{tseven}) becomes
\be
\d \qth = -\s \qth
\label{qone}
\ee
Thus in this equation there appears the quantity $S$ and not $\hat S$. Our goal is to get a closed system of equations in the hatted variables.
To this end we note that for the number unity we can write the hatted symbol
\be
\ih~\equiv~  e^{\int_{t_0}^t \p dt}\cdot 1~=~e^{\int_{t_0}^t \p dt}
\ee
Using $\hat I$ we can write (\ref{qone}) as 
\be
\d \qth = -{\sh\over\ih} ~\qth
\ee
Note that
\be
\d\ih = \ph
\ee
so we finally do have a closed system of equations in the hatted variables. We collect these equations together for later use
\bea
\d \ph&=&-K_1 \qth \label{qtwo}\\
\d \qth &=& -{\sh\over\ih} ~\qth \label{qthree}\\
\d \sh&=&K_2 \qth \label{qfour}\\
\d\ih &=& \ph \label{qfive}
\eea
Eq.(\ref{qsix}) for the $\gamma_i$ can also be written simply in hatted variables
\be
{d\over dt} \hat \gamma_k=-\delta_k \hat \qt
\label{qseven}
\ee
where
\be
\delta_k=\h [{1-W\over D-2}+w_k]
\label{qeight}
\ee

\subsection{Integrals of motion}

The hatted version of the basic equations allow us to note some simple integrals of the equations.

From (\ref{qtwo}) and (\ref{qfour}) we find immediately that
\be
\d~(K_2 \ph + K_1 \sh)=0
\ee 
which gives
\be
\sh=-{K_2\over K_1}~\ph +{\rm constant}
\ee
where the constant is determined by initial conditions.

From (\ref{qeight}) and (\ref{qtwo}) we find 
\be
\d \hat \gamma_k=-\delta_k \hat \qt = {\delta_k\over K_1} \d \ph
\label{mgk}
\ee
which gives
\be
\hat \gamma_k={\delta_k\over K_1} \ph + F_k
\label{ff}
\ee
where $F_k$ are constants determined by initial conditions.
Note that
\be
\sum_k {\delta_k\over K_1}=1
\ee
Since $\sum_k\hat \gamma_k=\hat P$, we see that we must have
\be
\sum_k F_k=0
\ee

\subsection{Physical ranges for parameters}

Note that 
\be
\p=\sum_i \gamma_i={d\over dt} ~\sum_i \log a_i={d\over dt} ~\log V={\dot V\over V}
\ee
where $V$ is the volume of the spatial torus.
We have to choose a direction of time to call positive, and we can use this freedom to require that  at the time $t_0$ where we give initial conditions
\be
\p(t=t_0)\ge 0
\label{ppos}
\ee

The integrating factors that converts un-hatted quantities to hatted ones is 
\be
\ih=e^{\int_{t_0}^t \p dt}={V\over V_0}
\label{ihfa}
\ee
Thus throughout the physical range of evolution we will have
\be
\ih> 0
\label{none}
\ee
From (\ref{defhat}) it follows that hatted and un-hatted variables have the same sign.

From (\ref{gtt}) we find that
\be
\h (\sum_i \gamma_i)^2-\h \sum_i \gamma_i^2=\h (\p^2- \q)=8\pi G \rho
\ee
so the energy density is
\be
\rho=-{1\over 16\pi G} \qt
\label{rhoq}
\ee
Our matter is made up of the quanta in string theory, and we have seen that the energy of different kinds of objects will be simply added
to obtain the total energy. Each of these quanta have a positive energy, so we will have 
\be
\rho > 0, ~~~~\qt < 0, ~~~\qth< 0
\label{oone}
\ee
The total energy is $E=\rho V$, so $\hat {\t Q}$ is just the total energy upto a (negative) constant
\be
\qth = {V\over V_0} \qt  = -{16\pi G\over V_0} ~E
\ee

We have found that $p_k=w_k \rho$. We will assume the dominant energy condition, which states that for each direction the pressure satisfies $p_k\le \rho$. (As in the case of energy density, the pressure is obtained by adding the pressure contributed by the different string theory objects in our state, and these satisfy the dominant energy condition.) Thus we have
\be
w_i \le 1
\ee 
for all $i$. This gives $W \le (D-1)$, or equivalently, $D-1-W \ge 0$. This implies that
\be
K_1={(D-1-W)\over 2(D-2)}\ge 0
\label{otwo}
\ee

From (\ref{oone}) and (\ref{otwo}) we find that the RHS of (\ref{qtwo}) is non-negative, so $\hat P$ is a non-decreasing function of time
\be
\d \ph \ge 0
\label{php}
\ee
From (\ref{ppos}) and (\ref{php}) we see that for all $t\ge t_0$ we will have
\be
\ph\ge 0
\label{ppos2}
\ee
From (\ref{qfive}) we see that 
\be
\d \ih \ge 0
\ee
so that $\ih ={V\over V_0}$ is a nondecreasing function of  time. Thus the Universe will not `recollapse'
to $V\r 0$. 

The variables $\s, \sh$ can have either sign, and this sign can change during the evolution.

\section{Solving the equations}
\setcounter{equation}{0}

We observe that in the system  (\ref{qtwo})-(\ref{qfive}), three of equations have $\qth$ on the right hand side. We can divide by $\qth$ and absorb it in the definition of time, by writing
\be
{1\over (-\qth)} ~{d\over dt} \equiv {d\over d\tau}
\ee
We have put in the negative sign because $\qth$ is negative; with this sign, the variable $\tau$ increases when $t$ increases.  Thus the $t, \tau$ variables are related by
\be
\tau=\int_{t_0}^t dt' ~(-\qth), ~~~~(t-t_0)=\int_0^\tau ~{d\tau'\over (-\qth)}
\label{time}
\ee
where we have chosen the lower limit of $t$ to be the time  $t_0$ where we specify initial conditions. The system (\ref{qtwo})-(\ref{qfive}) then gives
\bea
\du \ph &=& K_1\label{mone}\\
\du\qth &=& {\sh\over \ih} \label{mtwo}\\
\du\sh &=& -K_2\label{mthree}\\
\du\ih &=& -{\ph\over \qth}\label{mfour}
\eea
We can immediately solve (\ref{mone}) and (\ref{mthree}):
\bea
\ph&=&K_1\tau + A_1\label{mfive}\\
\sh&=&-K_2\tau + A_2\label{msix}
\eea
where $A_1, A_2$ are constants.
Now (\ref{mtwo}) and (\ref{mfour}) become
\bea
\du\qth&=& {\sh\over \ih}={(-K_2\tau + A_2)\over \ih}\label{mseven}\\
\du\ih&=&-{\ph\over \qth}= -{(K_1\tau + A_1)\over \qth}\label{meight}
\eea
From these equations we deduce that
\be
(\du \qth)\ih+\qth(\du \ih)= \du(\qth\ih)=-(K_1+K_2)\tau + (A_2-A_1)
\ee
which gives
\be
\qth\ih= -(K_1+K_2){\tau^2\over 2}+(A_2-A_1)\tau + A_3
\label{para}
\ee
where $A_3$ is another constant. Taking $\ih$ from this equation in substituting it in (\ref{mseven}) gives
\be
\du \qth =  { (-K_2 \tau + A_2)\over -(K_1+K_2){\tau^2\over 2}+(A_2-A_1)\tau + A_3} ~ \qth
\ee
or
\be
\du [\log (-\qth)]={ (-K_2 \tau + A_2)\over -(K_1+K_2){\tau^2\over 2}+(A_2-A_1)\tau + A_3}
\label{mlog}
\ee 
where we have written $(-\qth) $ in the argument of the $\log$ since $\qth$ is negative. The quadratic in the denominator on the right hand side can be written as
\be
-(K_1+K_2){\tau^2\over 2}+(A_2-A_1)\tau + A_3= -{(K_1+K_2)\over 2} ~(\tau-r_1) (\tau-r_2)
\label{roots}
\ee
where $r_1, r_2$ are the two roots of the quadratic. We now need to know if these roots are real or complex, and if they are real, then where $\tau$ lies on the real axis with respect to these roots. We will study these roots below, but for now  we write the formal solution to (\ref{mlog})
\be
(-\qth)= A_4 (\tau-r_1)^{-{2 (-r_1 K_2 + A_2)\over ( K_1+K_2)(r_1-r_2)}}(\tau-r_2)^{{2(-r_2 K_2 + A_2)\over (K_1+K_2) ( r_1-r_2)}}
\label{mqth}
\ee
where $A_4$ is a  constant.

Thus all variables $\ph, \qth, \sh, \ih$ have been expressed algebraically in terms of the time  parameter $\tau$. From (\ref{mfive})
we see that  upto a  suitable choice of origin and a constant scaling, $\tau$ is just the variable $\ph$. Thus if we use $\ph$ to measure time, then all other variables are given by rational functions of  this time. To get back to the physical problem however, we need to relate $\tau$ to $t$. This is done through (\ref{time})
\be
(t-t_0)={1\over A_4}\int_0^\tau (\tau'-r_1)^{{2 (-r_1 K_2 + A_2)\over ( K_1+K_2)(r_1-r_2)}}(\tau'-r_2)^{-{2(-r_2 K_2 + A_2)\over (K_1+K_2) ( r_1-r_2)}} d\tau'
\label{mintegral}
\ee
The integral on the RHS is given by an incomplete Beta function.  This function is defined by \cite{grad}
\be
B_x(p,q)=\int_0^x s^{p-1} (1-s)^{q-1} ds
\ee
The precise expression for (\ref{mintegral}) in terms of the incomplete Beta function will depend on the location of $\tau$ with respect to the roots $r_1, r_2$.

Since the relation between $t$ and $\tau$ is transcendental, we will analyze the solutions qualitatively to see the dynamical behavior that results for different choices of parameters.

\subsection{Different dynamical behaviors}

Consider the integral in (\ref{mintegral}). For what follows we recall eq.(\ref{php}) which says that  $\ph$ cannot decrease with time. There are three possible cases:

\bigskip

(a) The integral (\ref{mintegral}) diverges at a finite value of $\tau$. Then we reach $t=\infty$ with finite $\tau$. Then from (\ref{mfive}) we see that $\ph$ asymptotes to a finite constant. 
\bigskip

(b) The integral (\ref{mintegral}) diverges as $\tau\r\infty$. In this case $\ph\r\infty$ as $t\r\infty$. 

\bigskip

(c) The integral (\ref{mintegral}) converges. In this case we have a divergence $\ph\r\infty$ at a finite time $t$.

\bigskip

\subsection{Dependence on parameters}

We now wish to see which of the above behaviors results for which choices of parameters and initial conditions. Recall that   $\ih$ is positive (eq.(\ref{none})) and $\qth$ is negative (eq.(\ref{oone})). Thus the left hand side of (\ref{para}) is negative.  Consider the function
\be
f(\tau)\equiv (-\ih\qth) = {(K_1+K_2)\over 2}{\tau^2}-(A_2-A_1)\tau - A_3
\label{mf}
\ee
The physical values of parameters then requires 
\be
f\ge 0
\label{fpos}
\ee 
The function $f(\tau)$ describes a parabola. We have two cases:\footnote{Here and in other computations below we consider only generic values of the parameters for simplicity. For example we do not explicitly look at the border $K_1+K_2=0$ between the two cases below; such special cases can be easily worked out explicitly.} 

\begin{figure}[htbp] 
   \centering
\includegraphics[width=2in]{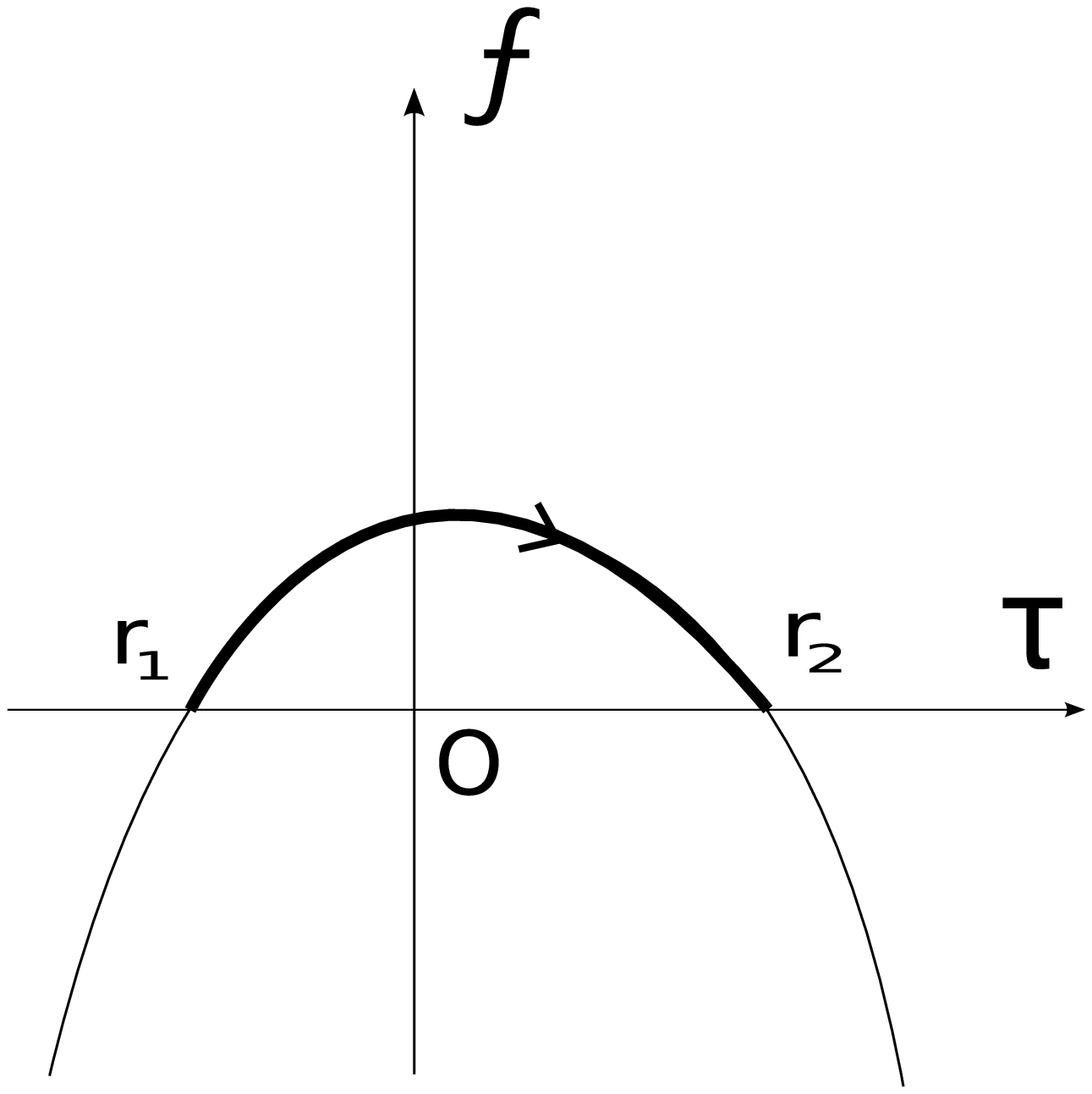}\hspace{1truecm}
   \includegraphics[width=2.3in]{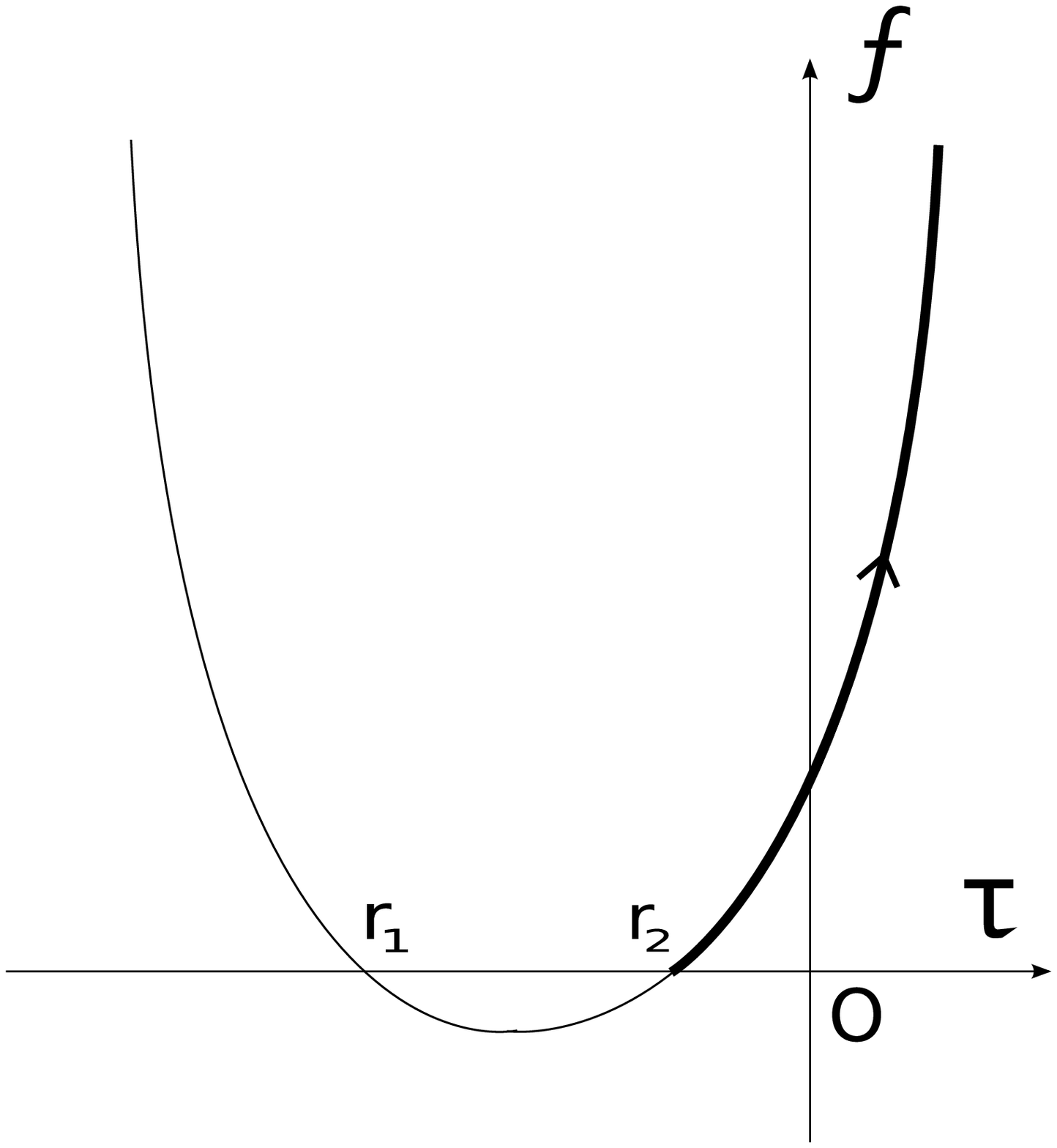}
   \vspace{.2truecm}
   \hspace{4.5truecm} (a) \hspace{7truecm} (b)
   \vspace{.5truecm}  
   \caption{(a) Downward facing parabola for $K_1+K_2<0$ \quad (b) Upward facing parabola for $K_1+K_2>0$. In each case a physical choice of parameters leads to motion along the bold line segment.}
   \label{parabolas}
\end{figure}

\subsubsection{$K_1+K_2<0$}

In this case the parabola is concave downwards. From (\ref{fpos}) we see that only the part of $f$ above the $\tau$-axis describes a physical evolution. If the roots $r_1, r_2$ in (\ref{roots}) are real, then the parabola will intersect the $\tau$-axis, and a part of the parabola will lie above this axis (Fig.\ref{parabolas}(a)). If the roots are complex, the parabola will be entirely below the $\tau$-axis.

In this case it is easy to show that $r_1, r_2$ will be real. The discriminant of the polynomial in (\ref{roots}) is
\be
\Delta=(A_2-A_1)^2+2 A_3(K_1+K_2)
\ee
Note that $A_3$ is the value of the negative quantity $\qth\ih$ at the initial time $\tau=0$, so
\be
A_3<0
\ee
Since $K_1+K_2<0$ in the present case, we find that all terms in $\Delta$ are positive and thus $\Delta\ge 0$. Thus the roots $r_1, r_2$ are real, and the parabola will look as in Fig.\ref{parabolas}(a). 
The evolution will take place on the solid part of the parabola in Fig.\ref{parabolas}(a). 
Since the evolution ends at a finite value of $\tau$, we find from (\ref{mfive}) that  $\ph$ asymptotes to a constant, and thus we will be in case (a).

The initial data is given at $\tau=0$, which must be a point between the two roots of the parabola. Thus $r_1<0$ and $r_2>0$. Since $r_1<\tau<r_2$ during the physical evolution the relation (\ref{mintegral}) should be written as
\be
(t-t_0)={1\over |A_4|}\int_0^\tau (\tau'-r_1)^{\alpha_1}(r_2-\tau')^{\alpha_2} d\tau'
\label{timea}
\ee
where we have defined
\be
\alpha_1={{2 (-r_1 K_2 + A_2)\over ( K_1+K_2)(r_1-r_2)}}, ~~~~\alpha_2={-{2(-r_2 K_2 + A_2)\over (K_1+K_2) ( r_1-r_2)}}
\ee
We find
\bea
(t-t_0)&=&{1\over |A_4|}(r_2-r_1)^{{K_1-K_2\over K_1+K_2}}~\left ( B_{\tau-r_1\over r_2-r_1}[\alpha_1+1, \alpha_2+1]~-~B_{-{r_1\over r_2-r_1}}[\alpha_1+1, \alpha_2+1] \right )\nn
\eea

\begin{figure}[htbp] 
   \centering
   \includegraphics[width=2.5in]{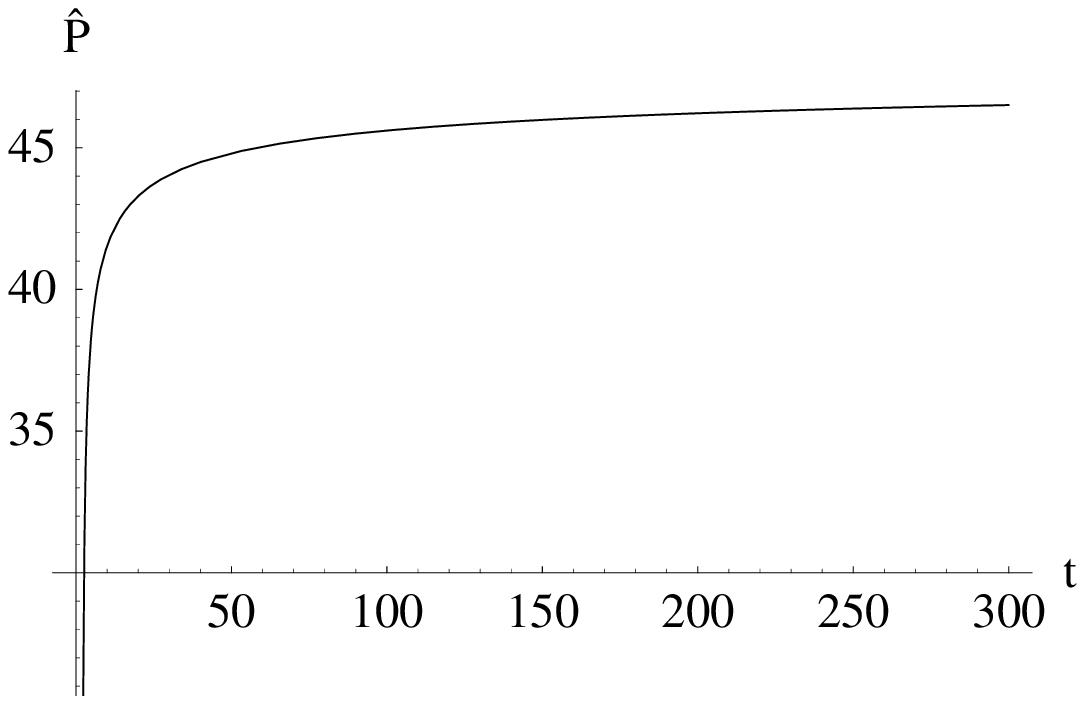} \hspace{1truecm}
    \includegraphics[width=2.5in]{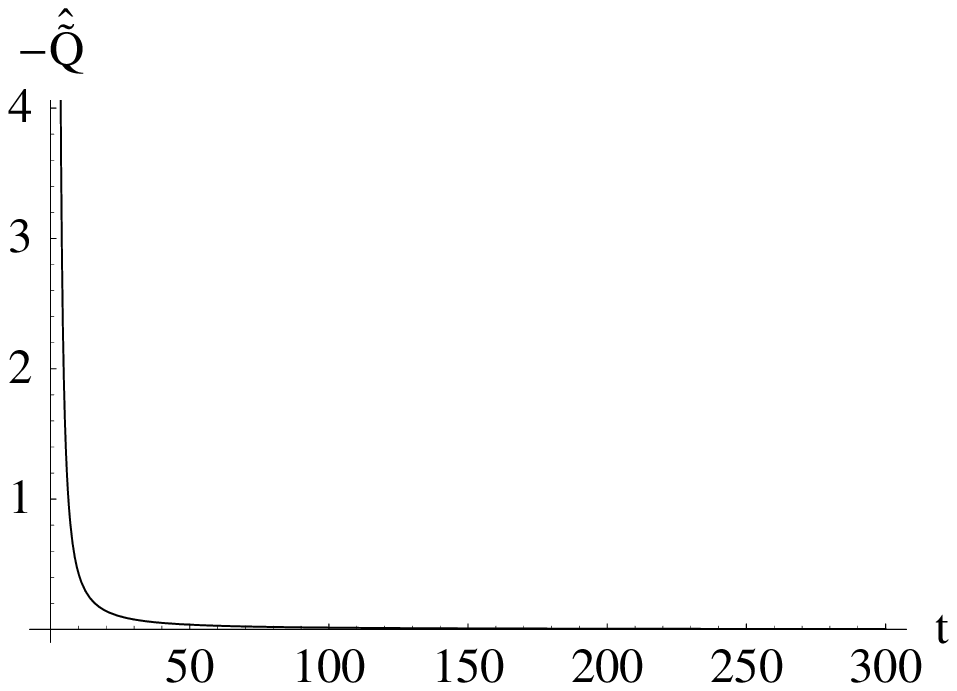}\\
\vspace{.5truecm}
    \includegraphics[width=2in]{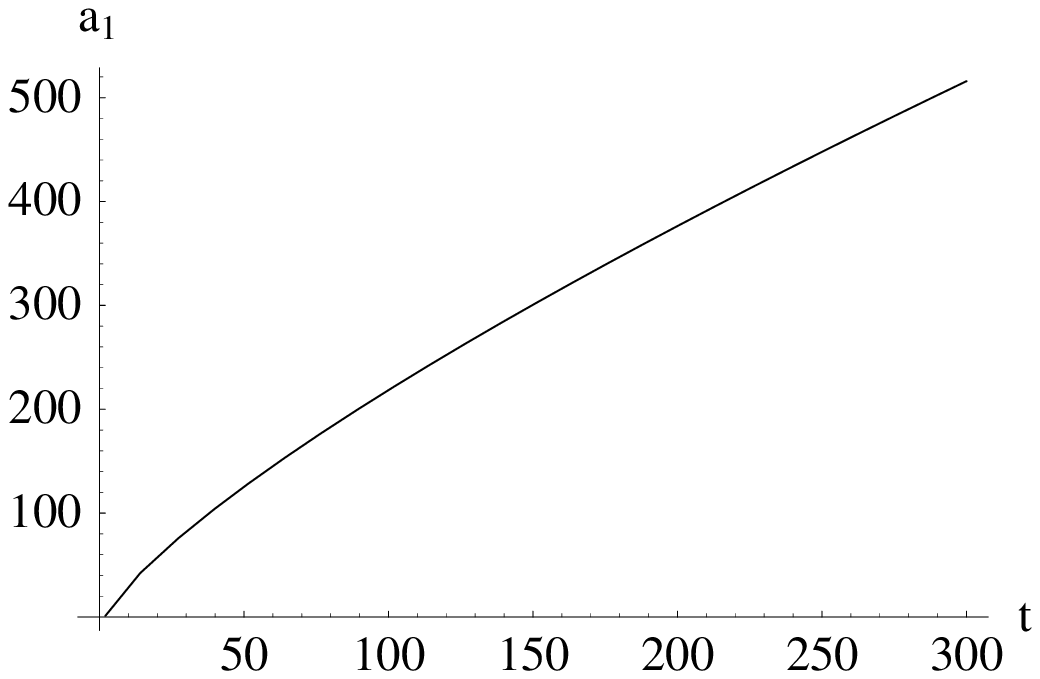}\hspace{.1truecm}
 \includegraphics[width=2in]{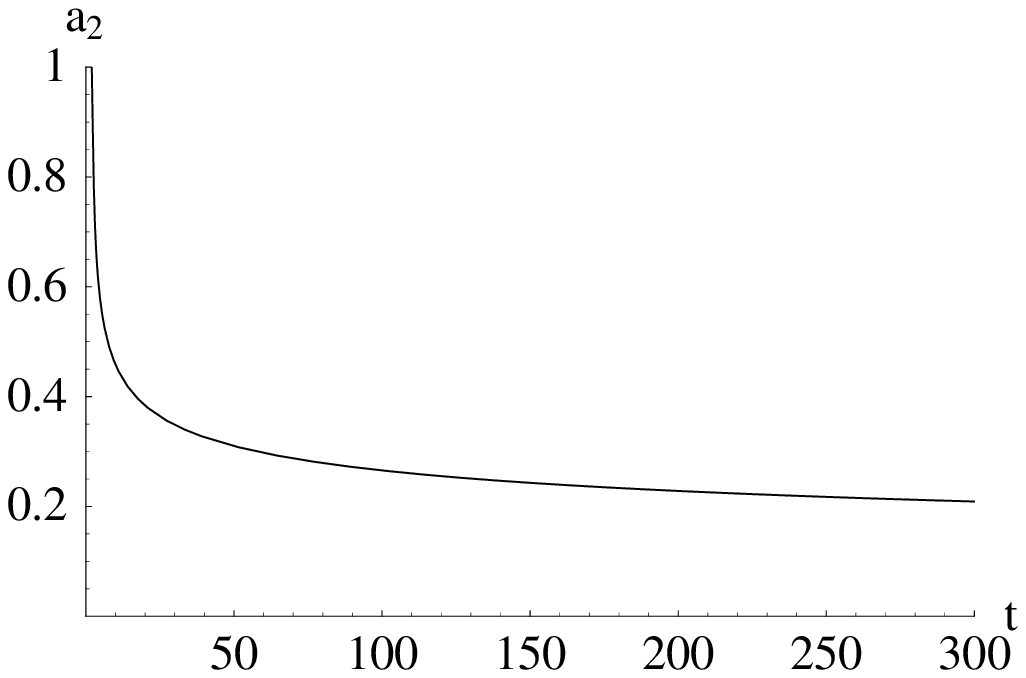} \hspace{.1truecm}
    \includegraphics[width=2in]{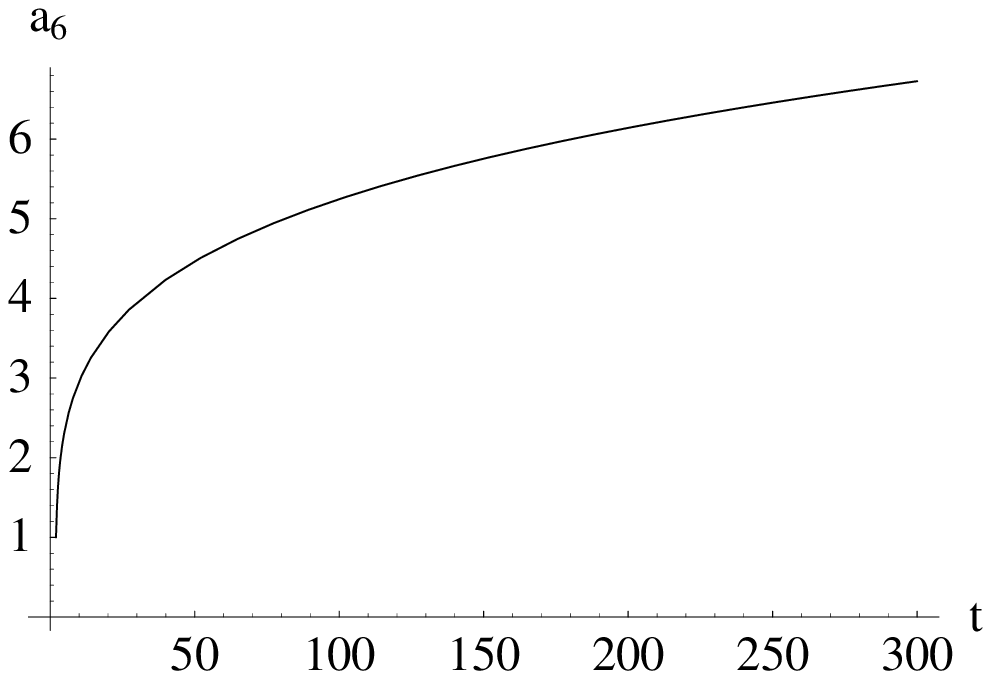} 
\vspace{.5truecm}
 \caption{Plots of $\ph, -\qth$ and a selection  of $a_i$ for $w_i=\{ .9, -.9, -.9, -.9, -.9, -.1, -.1, -.1, -.1, -.1\}$, a set that gives $K_1+K_2<0$ and illustrates case (a) behavior. We have taken $\gamma_i(2)=a_i(2)=1$ for all $i$.  We see that $\ph$ asymptotes to a constant.}
   \label{casea}
\end{figure}

In Fig.\ref{casea} we plot graphs for an example that illustrates case (a). We let  $w_i=\{ .9, -.9, -.9, -.9, -.9, -.1, -.1, -.1, -.1, -.1\}$. We have set  $t_0=2$ and have taken  $\gamma_i(t=t_0)=1$, $a_i(t=t_0)=1$ for all $i$. We plot $\ph, -\qth$, and one $a_i$ from each set having the same $w_i$.  Note that $-\qth$ is proportional to the total energy in the Universe.

\subsubsection{$K_1+K_2>0$}

In this case the parabola is concave upwards. With a little more effort we can again show that for physically allowed initial conditions, the discriminant $\Delta$ is positive, and thus $r_1, r_2$ are real; this computation is done in Appendix (\ref{aa1}). Thus the parabola intersects the $\tau$-axis, as shown in Fig.\ref{parabolas}(b). From (\ref{fpos}) we see that physical motion can take place only on the two segments of the parabola above the $\tau$-axis. Thus there appear to be two possible branches, a `left' branch and a `right' branch. With some effort we can  show that for a physical choice of parameters, we cannot be on the left branch; this is done in Appendix (\ref{aa2}).
Thus motion takes place only on the right branch, as indicated by the solid part of the parabola.

We see that the motion will extend to $\tau\r\infty$; thus from (\ref{mfive}) we see that we will be in case (b) or in case (c). To distinguish between these cases consider the integral (\ref{mintegral}). Since we are on the right branch above the $\tau$-axis we see that $(\tau-r_1)>0$ and $(\tau-r_2)>0$. Thus from (\ref{mqth}) we find $A_4$ to be a real positive constant. In (\ref{mintegral}) we find that the large $\tau$ behavior gives
\be
t-t_0\sim  \int^\tau d\tau' (\tau')^{-{2K_2\over K_1+K_2}}\sim {(\tau)^{-{2K_2\over K_1+K_2}+1}}
\label{timeb}
\ee

Thus the convergence of this integral is determined by the sign of $\mu\equiv -{2K_2\over K_1+K_2}+1$. Note that we have $K_1+K_2>0$ in the present part of the analysis. So we can equally well ask for the sign of $(K_1+K_2)\mu=K_1-K_2$. We thus have the two cases

\bigskip

(i) $K_1-K_2>0$:\quad In this case the integral (\ref{mintegral}) diverges at $\tau=\infty$, and we have case (b).

\bigskip

(ii) $K_1-K_2<0$:\quad In this case the integral (\ref{mintegral}) converges and we have case (c). We note however in Appendix (\ref{aa3}) that we can have this case only if at least one of the $w_i$ is less than $-1$. It is conventionally assumed that $w_i$ lie in the range $-1\le w_i\le 1$. The upper limit comes from the dominant energy condition, but there is no strong reason to require the lower limit. For the quanta that we get from string theory though we do have $-1\le w_1\le 1$, as can be seen from the definition (\ref{wnc}).

\bigskip

In either of these cases (b),(c) 
we are on the `right branch' of the parabola in Fig.\ref{parabolas}(b). Thus the point $\tau=0$ where the initial data is specified  lies to right of the two roots of the
parabola. So $r_1<r_2<0$, and we find from (\ref{mintegral})
\be
(t-t_0)={(r_2-r_1)^{\alpha_1+\alpha_2+1}\over |A_4|}
\left (
B_{\tau-r_2\over \tau-r_1}[\alpha_2+1, -\alpha_1-\alpha_2-1)]-
B_{r_2\over r_1}[\alpha_2+1, -\alpha_1-\alpha_2-1)] \right )\nn
\ee

\begin{figure}[htbp] 
   \centering
   \includegraphics[width=2in]{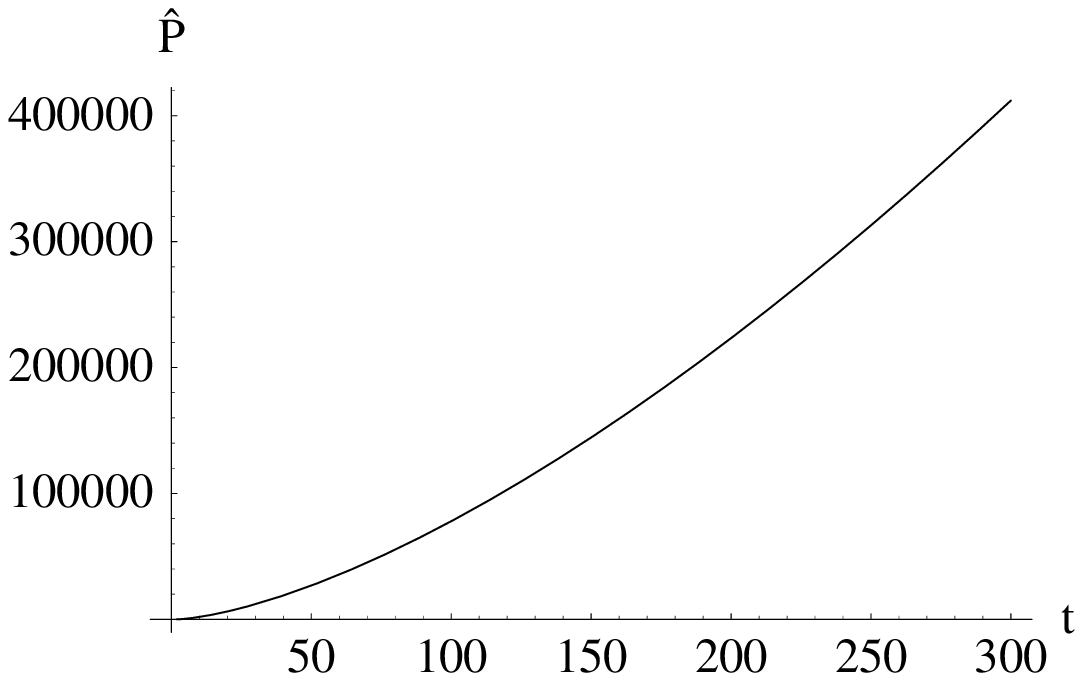} \hspace{.1truecm}
    \includegraphics[width=2in]{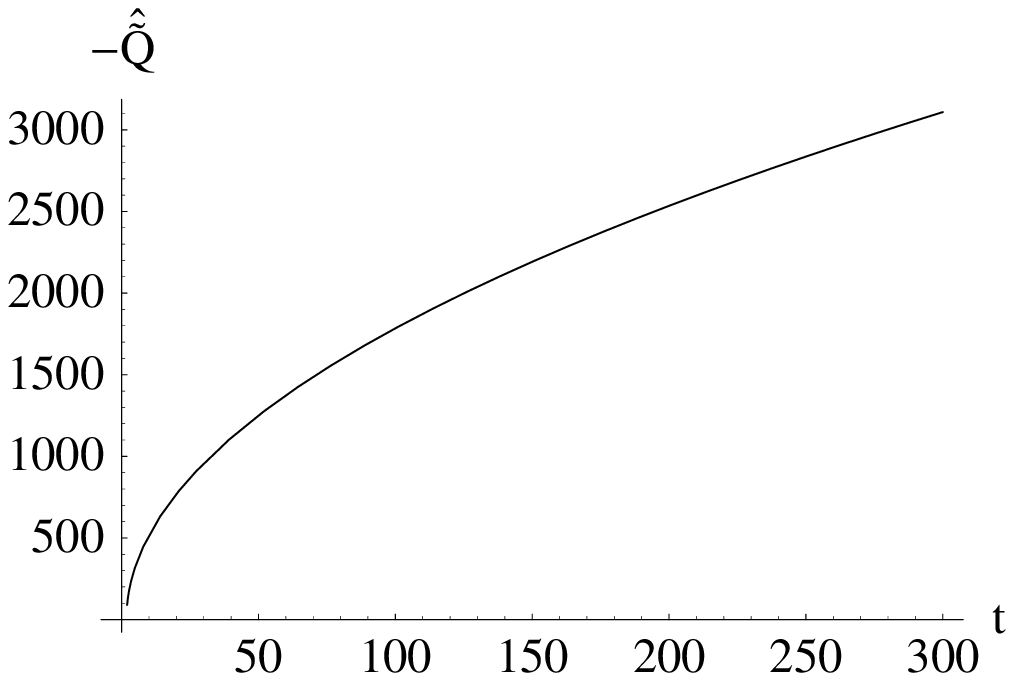} \hspace{.1truecm}\includegraphics[width=2in]{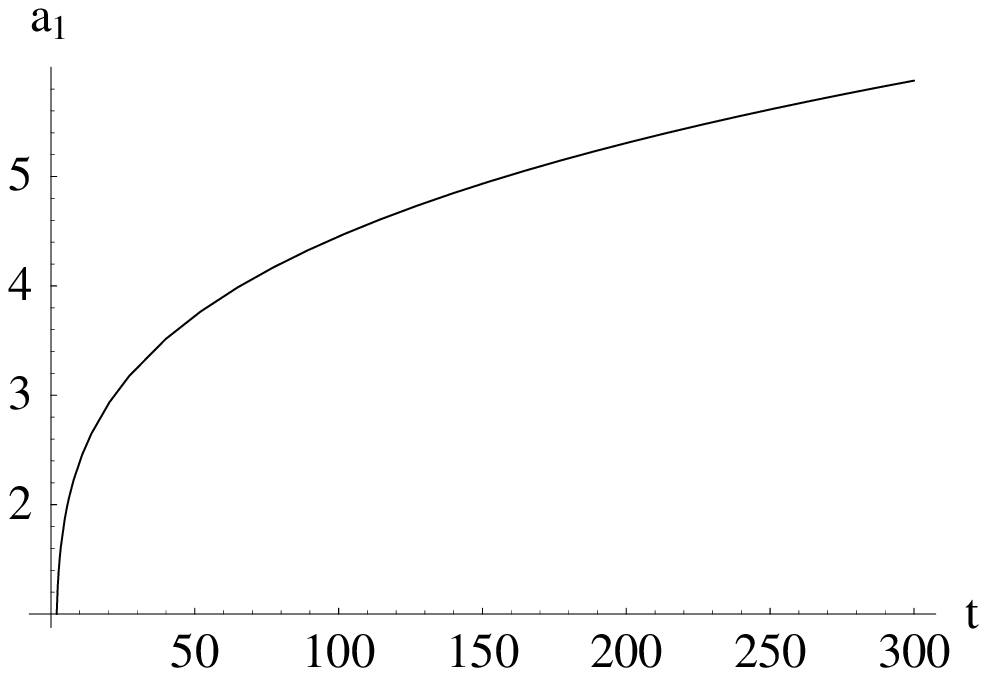}
\vspace{.5truecm}
 \caption{Plots of $\ph, \qth, a_1$ for the choice $w_i=-.2$ for all $i$. This gives $K_1+K_2>0$, $K_1>K_2$, and thus case (b) behavior. $\ph$ grows without bound. }
   \label{caseb}
\end{figure}

In Fig.\ref{caseb} we plot graphs for an example that illustrates case (b). We have taken $w_i=-.2$ for all $i$.  We have set  $t_0=2$ and have taken  $\gamma_i(t=t_0)=1$, $a_i(t=t_0)=1$ for all $i$. We plot $\ph, -\qth$, and $a_1$.  

\subsection{Solving for $\gamma_i, a_i$}

From (\ref{mgk}) we find
\be
(-{1\over \qth}) ~ \d \hat\gamma_k =\du\hat\gamma_k= \delta_k
\ee
which gives
\be
\hat\gamma_k=\delta_k\tau+f_k
\label{forapp2}
\ee
where $f_k$ are constants. Since
\be
\sum_k\hat\gamma_k=\ih\sum_k\gamma_k=\ih P=\ph
\ee
we have one relation between  the $f_k$
\be
\sum_k f_k = \ph(\tau=0)=A_1
\ee

Now note that
\be
\d (\log a_k) ={\ad_k\over a_k}=\gamma_k={\hat\gamma_k\over \ih}
={\delta_k\tau+f_k\over \ih}
\ee
Thus
\be
\du (\log a_k)=(-{1\over \qth}) ~ \d (\log a_k)=-{(\delta_k\tau+f_k)\over \qth\ih}= -[ {(\delta_k\tau+ f_k)\over -{(K_1+K_2)\over 2} ~(\tau-r_1) (\tau-r_2)}]
\ee
where we have used (\ref{para}),(\ref{roots}). This gives
\be
a_k=C_k ~(\tau-r_1)^{{2(\delta_k r_1 + f_k)\over (K_1+K_2)(r_1-r_2)}}(\tau-r_2)^{-{2(\delta_k r_2+f_k)\over (K_1+K_2)(r_1-r_2)}}
\label{aevolve}
\ee
where $C_k$ are constants.

\subsection{Evolution as a function of $t$}

We have seen above that all variables can be expressed as algebraic functions in terms of the time parameter $\tau$. But in the metric (\ref{metric}) the natural parameter is $t$, and the relation between $\tau$ and $t$ was given through the incomplete Beta function, so this relation is not easy to picture. We can however find that evolution as a function of $t$ at late times, where the relation between $\tau$ and $t$ simplifies.

\subsubsection{Case (a) at large $t$}

From Fig.\ref{parabolas}(a) we see that the evolution takes us towards  the root $\tau=r_2$. In Appendix (\ref{aa4}) we show that 
\be
\alpha_2+1<0
\label{alpharel}
\ee
Then we see from (\ref{timea}) that $t$ diverges as $\tau\r r_2$
\be
t\sim (r_2-\tau)^{\alpha_2+1}
\ee
and from (\ref{aevolve}) we get
\be
a_k\sim t^{{2(\delta_k r_2+f_k)\over 2(-r_2 K_2 +A_2)-(K_1+K_2)(r_1-r_2)}}
\label{powert}
\ee
Thus the late time behavior of the  $a_k$ depends on the initial conditions that specify parameters like $f_k$ and $A_1,A_2,A_3$ which give the roots $r_1, r_2$. In Appendix (\ref{aa4}) we observe that the power of $t$ in (\ref{powert}) can be written in a simpler form \be
a_k\sim t^{{\gamma_k(\tau_2)\over \p(\tau_2)}}
\label{forapp}
\ee

\subsubsection{Case (b) at large $t$}

In this case we have $t\r\infty$ as $\tau\r\infty$. Let us examine this large $t$ behavior. From (\ref{timeb}) we get
\be
\tau \sim  t^{{K_1+K_2\over K_1-K_2}}
\ee
From (\ref{aevolve}) we find
\be
a_k \sim t^{{2\delta_k\over K_1-K_2}}\sim t^{\beta_k}
\ee
where we have used the expression  (\ref{qeight}) for $\delta_k$ and the expressions (\ref{k1}),(\ref{k2}) for $K_1, K_2$ to recognize that the power of $t$ is the same as the power $\beta_k$ appearing in (\ref{betak}) for the
Kasner type solution. Thus we see that  evolution for case (b) asymptotes at late times to the Kasner type solution for the given $w_k$.\footnote{In \cite{greene2} it was also observed (from numerical computations) that there was a kind of `attractor mechanism', so that solutions with generic initial data became similar to each other at late times.}

\subsubsection{Case (c) for $t\r t_f$}

In this case the $\tau$ integral converges
\be
{1\over |A_4|}\int_0^\infty (\tau-r_1)^{{2 (-r_1 K_2 + A_2)\over ( K_1+K_2)(r_1-r_2)}}(\tau-r_2)^{-{2(-r_2 K_2 + A_2)\over (K_1+K_2) ( r_1-r_2)}}\equiv t_f-t_0
\ee
To find the behavior as $t\r t_f$ we note from (\ref{mintegral}) that in this limit $\tau$ is large and we have
\be
\tau\sim (t_f-t)^{K_1+K_2\over K_1-K_2}
\ee
From (\ref{aevolve}) we get
\be
a_k\sim (t_f-t)^{{2\delta_k\over K_1-K_2}} \sim (t_f-t)^{\beta_k}
\ee
where we have noted that the powers of $(t_f-t)$ appearing here are the same as those that appear in the Kasner type power law solution $a_k\sim t^\beta_k$.

\section{Discussion}
\setcounter{equation}{0}

We have assumed an equation of state suggested by black holes; the energy goes to creating  `fractional branes' which have a high entropy and can thus  dominate over other kinds of matter.  If we assume that the state of the Universe is always the maximal entropy state for the given energy $E$, then we get an equation of state $p_i=w_i\rho$, for which the dynamic can be solved analytically. 

We discussed some of the ideas behind the fractional brane state in section (\ref{branesec}). If  we have a few branes in a given volume of space, then these branes will tend to annihilate to massless quanta. But if we increase the energy $E$ in the given volume to very large values, then it becomes entropically favorable to produce a large number of suitable sets of branes and anti-branes. Branes in such a set are mutually BPS, and `fractionate' each other, producing an entropy that grows more rapidly with $E$ than the energy of radiation or a Hagedorn type string or brane gas. As we discussed in section (\ref{branesec}),  black hole results suggest that at these densities the fractional brane quanta seem to behave as essentially free objects. Thus the energy $E$ and pressures $p_i$ for the state are given by just adding the contributions from  the branes and antibranes. The fractional branes do not find it easy to find each other and annihilate; the rate of this annihilation
can be computed for black holes where it reproduces exactly the rate of Hawking radiation.

If we follow the state of the present Universe backwards in time, the energy density keeps increasing. It has been postulated that for sufficiently early times we reach a Hagedorn phase of strings. This phase has no pressure,  but does have energy; thus as we go further back in time $E$ does not change ($dE=-PdV$) but $\ad\ne 0$, so the density grows still further. This suggests that at sufficiently early times the fractional brane state should be present.

We have not analyzed the astrophysical implications of the evolution we find;   there are many issues to be addressed  and we hope to carry out a detailed study of these elsewhere. Here we outline some of these issues and raise some relevant questions.

We have not taken any specific choice of the $w_i$; rather we solved the problem for an arbitrary set of $w_i$. Which choice of $w_i$ that gives the largest possible entropy for M theory on $T^{10}$? We can find large sets of mutually BPS branes with 10 compact directions, but we need to prove that choosing branes and antibranes of these varieties will indeed give the entropy (\ref{entropyass}). We thus need to generalize the brane constructions that have worked so well for black holes and  understand the entropy of high density states in string theory.

For generic choices of $w_i$ we get a power law expansion, and for many choices the  power  is too low to give us some kind of `inflation'.  Note than in inflation we find the inflaton in a low entropy state, while in  string/brane gas approach we seek the {\it maximal} entropy state for the given energy. In this sense we are closer to the latter approach; we look for a maximal entropy state but observe that the Hagedorn gas is not the highest entropy state in string theory.

Inflation gives a reason for the sky to look homogenous on large length scales. But there is a different source of long distance correlations that can arise with the fractional brane gas. In black holes semiclassical analysis suggests that quantum gravity effects are confined to planck or string length, but because of fractionation we find that the brane bound state has a size that grows with the number of quanta in the state, and thus we get nonlocal quantum effects all across the interior of the horizon \cite{review1}. So in our Cosmological fractional brane state we can   also expect quantum correlations across macroscopic distances.

What is the fate of the fractional branes as the Universe expands? 
Consider the 3-charge entropy (\ref{el}) and the 4-charge entropy (\ref{eight}). For a given energy $E$, is it more advantageous to create three kinds of charges or four? We see that if all the $n_i$ exceed unity, then the 4-charge entropy is higher, but if one of them drops below unity, then the 3-charge entropy will be higher. Such a transition was studied in \cite{emission,cgm} where it gave a microscopic description of the black hole -- black string transition. In our present problem the mass of each type of brane changes as the $a_i$ evolve, and it is possible that some $n_i$ drops below unity. In that case we would have to continue the evolution with a different set of  branes and thus a different set of $w_i$.

Clearly it is crucial to know how many branes form the fractional brane state; this determines the $n_i$. Thus if the Universe was infinite and all branes were linked up to form the fractional brane state, then each $n_i$ would be infinite and the entropy per unit volume would diverge. (Note that doubling the volume more than doubles the entropy if three or more types of charges are involved.) It is clearly entropically advantageous for more and more of the matter to be linked up in the fractional brane state, but when the Universe starts the islands of matter so linked are presumably small; they might then grow rapidly as different islands come into causal contact.

Note that the {\it density} of matter in a fractional brane state need not be high as long as enough {\it total} matter is linked up in the state.  For example we can make a black hole with arbitrarily low density matter as long as there is enough total energy $E$ in the ball of matter. Any fractional brane matter left over today could show up as a dark component with its own dynamics. 

All these are interesting questions, and we regard the present work as just a first pass on the problem with these ideas; we hope to return to the above issues elsewhere.

\section*{Acknowledgements}

We thank S. Das, R. Furnstahl,  S. Giusto, D. Kabat and J. Michelson for many helpful comments.
This work was supported in part by DOE grant DE-FG02-91ER-40690.

\appendix
\section{Conditions arising from physical ranges of parameters}
\label{first}
\renewcommand{\theequation}{A.\arabic{equation}}
\setcounter{equation}{0}

First we collect together some definitions and relations that will be used to establish the inequalities in later sections of the Appendix.

Our spacetime dimension is $D$; thus the number of space directions is
\be
d\equiv D-1
\ee
We can regard the $w_i$ as a vector in $d$ dimensional space
\be
\vec w=\{w_1, w_2, \dots , w_d\}
\ee
Similarly
\be
\vec\gamma=\{\gamma_1, \gamma_2, \dots , \gamma_d\}
\ee
It will be convenient to define another $d$-dimensional vector which has all components unity
\be
\vec 1= \{ 1,1,\dots , 1\}
\ee
Then we find
\be
W=\vec 1\cdot \vec w, ~~~~U=\vec w\cdot \vec w
\ee
and
\be
\p=\vec \gamma\cdot \vec 1, ~~~~\q=\vec\gamma\cdot \vec\gamma, ~~~~\s=\gamma\cdot \vec w
\label{pqs}
\ee
Note that
\be
|\vec w|=\sqrt{U}, ~~~~|\vec 1|=\sqrt{d}
\ee
Let the angle between $\vec 1$ and $\vec w$ be $\theta$. 

Let the initial value of $\vec \gamma$ be $\vec \gamma_0$. The vectors $\vec 1$ and $\vec w$ define a plane in  $d$ dimensional space. We can decompose $\vec\gamma_0$ into a part along this plane and a part perpendicular to this plane
\be
\vec\gamma_0=\vec\gamma_{\parallel}+\vec\gamma_\perp
\ee
Let the angle between $\vec \gamma_\parallel$ and $\vec 1$ be $\phi$.

Note that the equations (\ref{qtwo})-(\ref{qfive}) are invariant under the rescaling
\be
\ph\r\mu\ph, ~~~~\qth\r\mu^2\qth, ~~~~\sh\r\mu\sh, ~~~~t\r\mu^{-1}t
\ee
where $\mu$ is a constant.
This rescaling just corresponds to scaling time, which scales  the $\ad_i$ while keeping fixed the $a_i$. Thus the initial $\gamma_i$ can be scaled without changing the essential behavior of the evolution. We will use this freedom to set  
\be
|\vec\gamma_\parallel|=1
\label{gpara}
\ee
This will simplify our expressions without loss of generality. We have
\be
(\vec 1\cdot\vec \gamma_0)=(\vec 1\cdot \vec \gamma_\parallel)=\sqrt{d}\cos\phi, ~~~~
(\vec w \cdot \vec\gamma_0)=(\vec w\cdot \vec \gamma_\parallel)=\sqrt{V}\cos(\theta-\phi), ~~~~(\vec \gamma_0\cdot\vec\gamma_0)=1+|\gamma_\perp|^2
\label{asix}
\ee

With a little algebra we find from the definitions (\ref{k1}),(\ref{k2})
\bea
K_1+K_2&=&-{1\over 2 (D-2)} \left ( [(D-1) U - W^2]-[U+(D-1)-2W]\right )\nn
&=&-{1\over 2 (d-1)} \left ( [ Ud-W^2]-[U+d-2W]\right)\nn
&=&-{1\over 2(d-1)}\left ( [(\vec 1\cdot \vec 1)(\vec w\cdot \vec w)-(\vec 1\cdot \vec w)^2] - [(\vec 1\cdot \vec 1)+ (\vec w\cdot \vec w)-2(\vec 1\cdot \vec w)]\right )\nn\label{aeightp}\\
&=&-{1\over 2(d-1)} \left ( [ Ud\sin^2\theta ] - [ U+d-2\sqrt{Ud}\cos\theta] \right )
\label{aeight}
\eea

Using (\ref{mtwo}),(\ref{mfour}) we get for the function $f$ giving the parabola (\ref{mf})
\be
\du f(\tau) = -\left (  \ih \du \qth + \qth \du \ih \right ) =\ph-\sh
\label{qsrel}
\ee

\subsection{Proof that $\Delta\ge 0$ for the case $K_1+K_2>0$}\label{aa1}

Recall that $\vec\gamma_0=\vec\gamma_\parallel+\vec\gamma_\perp$. We can write
\be
\vec\gamma_\parallel=\alpha \vec 1+\beta\vec w
\label{sspre}
\ee
At the initial time $t=t_0$ we see from  (\ref{defhat})  that hatted  variables equal the corresponding un-hatted variables, and from (\ref{ihfa}) we note that $\hat I(t_0)=1$. Thus using (\ref{asix})
\bea
(\vec\gamma_0\cdot\vec 1) &=& \p(\tau=0)=\ph(\tau=0)=A_1\label{ssone}\\
(\vec\gamma_0\cdot \vec w)&=&\s(\tau=0)=\sh(\tau=0)=A_2 \label{sstwo}
\eea
Taking the inner product of  (\ref{sspre}) with $\vec 1, \vec w$ we get two equations, solving which we get
\be
\alpha={ A_1 U - A_2 W\over Ud-W^2} , ~~~~~~\beta={-A_1 W+A_2 d\over U d-W^2} 
\ee
Note further that
\be
(\vec\gamma_0\cdot\vec\gamma_0)=\q(\tau=0)=(\qt+\p^2)(\tau=0)=(\qth+\ph^2) (\tau=0)= A_3+A_1^2
\ee
We now compute
\bea
|\vec\gamma_\perp|^2&=&|\vec\gamma_0-\alpha \vec 1-\beta\vec w|^2\nn
&=&A_3+{A_1^2 \left (U(d-1)-W^2 \right ) - A_2^2 d + 2 A_1 A_2 W\over Ud-W^2}
\eea
Since $|\vec\gamma_\perp|^2\ge 0$,  we get
\be
A_3\ge - [{A_1^2 \left (U(d-1)-W^2 \right ) - A_2^2 d + 2 A_1 A_2 W\over Ud-W^2}]
\ee
Note that in the present case we have $K_1+K_2>0$, so we can multiply the above inequality by $K_1+K_2$ without reversing the sign of the inequality. We thus have
\bea
\Delta &=& (A_1-A_2)^2+2 A_3(K_1+K_2)\nn
&\ge & (A_1-A_2)^2 -2(K_1+K_2){A_1^2 \left (U(d-1)-W^2 \right ) - A_2^2 d + 2 A_1 A_2 W\over Ud-W^2}\nn
&=& {[~A_1 \left (U(d-1)+W(1-W)\right ) +A_2 \left ( W-d\right ) ~] ^2   \over   (d-1)(Ud-W^2)  }
\eea
Note that 
\be
(Ud-W^2)= (\vec w\cdot \vec w)(\vec 1\cdot \vec 1) - (\vec 1\cdot \vec w)^2 \ge 0
\ee
by the Schwartz inequality. Thus we have
\be
\Delta\ge 0
\ee

\subsection{Absence of left branch of parabola for $K_1+K_2>0$}\label{aa2}

In Fig.\ref{parabolas}(b) positivity of energy density requires that we be above the $\tau$-axis.  Above this axis we find a left branch and a right branch of the parabola, and we wish to show that the left branch is not allowed for a physical choice of parameters.

\begin{figure}[htbp] 
   \centering
   \includegraphics[width=5in]{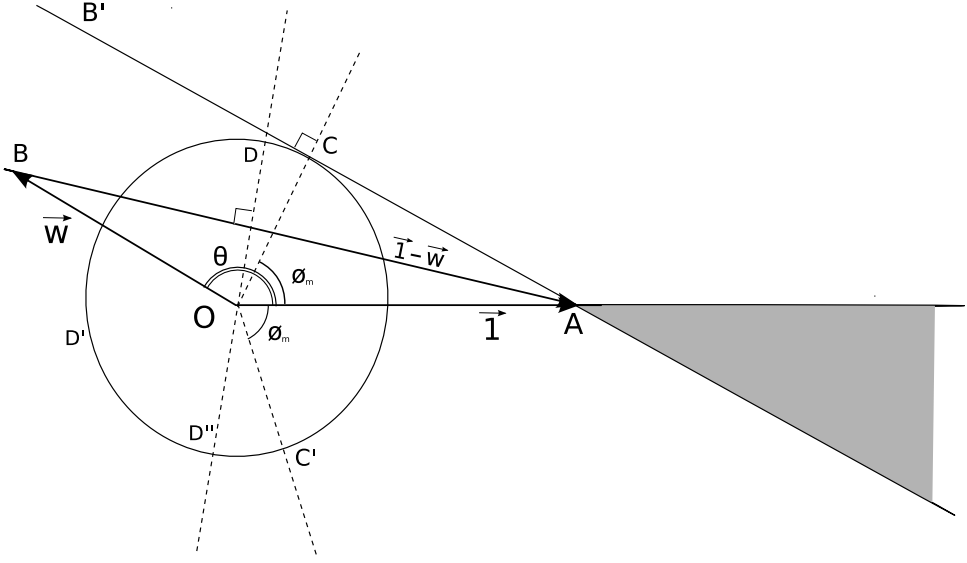} 
   \caption{Graphical representation of physical constraints for $K_1+K_2>0$, left branch of parabola.}
   \label{figpara}
\end{figure}

We will use four constraints that together will rule out the left branch:

\bigskip

(i) \quad In Fig.\ref{figpara}(a) we have drawn the vector $\vec 1$ as  OA, and the vector $\vec w$ as OB.  The angle between these vectors is $\theta$. Let $L$ be the length of BA. Let $L_\perp$ be the length of the perpendicular from O to the line BA. 

Note that in the present case we have $K_1+K_2>0$. Using (\ref{aeightp}) this gives
\be
[(\vec 1\cdot \vec 1)(\vec w\cdot \vec w)-(\vec 1\cdot \vec w)^2] < [(\vec 1\cdot \vec 1)+ (\vec w\cdot \vec w)-2(\vec 1\cdot \vec w)]
\label{ineqb}
\ee
Let the area of the triangle BOA be $A_\triangle$. The left side of  is (\ref{ineqb}) is
\be
(\vec 1\cdot \vec 1)(\vec w\cdot \vec w)-(\vec 1\cdot \vec w)^2=(\sqrt{U}\sqrt{d}\sin\theta)^2= (2A_\triangle)^2=(LL_\perp)^2
\ee
The right hand side of (\ref{ineqb}) is $|(\vec 1-\vec w)|^2=L^2$. Thus (\ref{ineqb}) gives us the constraint
\be
L_\perp<1
\label{lperpeqn}
\ee

\bigskip

(ii) \quad From the dominant energy condition we have $w_i\le 1$ for all $i$. This gives $W=\sum_i w_i \le d$, so
\be
d-W=(\vec 1-\vec w)\cdot \vec 1 \ge 0
\label{vecvec}
\ee

\bigskip

(iii) \quad  From (\ref{rhoq}) and the positivity of the energy density $\rho$ we have at $\tau=0$
\be
\qt=\q-\p^2 <0
\ee
From (\ref{pqs}) and using (\ref{gpara}) we get
\be
1+|\vec\gamma_\perp|^2-\p^2(\tau=0)<0
\ee
which gives 
\be
\p(\tau=0)>1
\label{ppp}
\ee

\bigskip

(iv) \quad Since we are looking at the left branch of the parabola given by the function $f(\tau)$, we have $\du f<0$. Using (\ref{qsrel}) we find
\be
\sh > \ph
\label{sp}
\ee

\bigskip

Now we will use these four constraints. Since $|\vec\gamma_\parallel|=1$, we draw a unit circle with center O; the tip of $\vec\gamma_\parallel$ must lie on this circle. We draw the tangent AB' to this circle, touching the circle at C. 

\bigskip

(i') \quad The lines AO and AB' form a wedge. There are two parts of this wedge: one to the left of $A$ and one to the right; the latter is shaded. From simple geometry using the condition (i)  (eq.(\ref{lperpeqn})) we find that the point B must lie inside this wedge. 

\bigskip

(ii') \quad The condition (ii) tells us that B cannot lie in the shaded part of the wedge. This follows because the line BA is the vector $\vec 1-\vec w$ and OA is the vector $\vec 1$. For B in the shaded part of the wedge the angle between BA and OA is obtuse, and thus   (\ref{vecvec}) is violated.

\bigskip

(iii') Recall that the angle between $\vec \gamma_\parallel$ and $\vec 1$ is $\phi$. The condition (\ref{ppp}) gives 
\be
\p=(\vec 1\cdot \vec \gamma_0)=(\vec 1\cdot \vec\gamma_\parallel)=\sqrt{d}\cos\phi>1
\ee
This tells us that 
\be
|\phi|<\phi_m\equiv \cos^{-1} {1\over \sqrt{d}}
\ee
Thus the tip of $\gamma_\parallel$ must lie in the arc CC'. 

\bigskip

(iv') \quad The condition (\ref{sp}) is
\be
(\vec 1-\vec w)\cdot \vec \gamma=(\vec 1-\vec w)\cdot \vec \gamma_\parallel<0
\ee
This tells us that the tip of $\vec \gamma_\parallel$ must lie in the arc DD'D''.

\bigskip

Finally, simple geometry tells us that if B is indeed  in the wedge indicated then
the arcs CC' and DD'D'' do not overlap. This proves that the four constraints (i)-(iv) are incompatible, and the left branch of the parabola is not obtained for physical values of parameters. 

\subsection{Impossibility of case (c) for $-1< w_i\le 1$}\label{aa3}

From the Schwartz inequality we have
\be
(\vec 1\cdot \vec 1) (\vec w\cdot \vec w) - (\vec 1\cdot \vec w)^2\ge 0
\ee
which gives
\be
U- {W^2\over d}\ge 0
\label{aone}
\ee
A little algebra gives
\be
U-{W^2\over d}=2(-K_1-K_2+2{d-1\over d}~K_1^2 )
\ee
Using (\ref{aone}) we find
\be
K_2\le -K_1+2{d-1\over d}~K_1^2
\label{atwo}
\ee
Recall from (\ref{otwo}) that $K_1>0$; this followed from the dominant energy condition ${p_i/ \rho}=w_i\le 1$ which we have assumed. 
Thus we can divide by $K_1$ in  the above inequality without reversing the inequality signs
\be
{K_2\over K_1}\le -1+2{d-1\over d}~K_1
\label{athree}
\ee
Now suppose that we have
\be
w_i>-1, ~~~~~i, 1, \dots d
\label{acondition}
\ee
Then
\be
W=\sum_i w_i > -d
\ee
Then from the expression (\ref{k1}) for $K_1$ we find
\be
K_1={D-1-W\over 2(D-2)}={d-W\over 2(d-1)}< {d\over d-1}
\ee
Using this in (\ref{athree}) we have
\be
{K_2\over K_1}<1 
\ee
Noting again from  (\ref{otwo}) that $K_1>0$, we can multiply through in this inequality without reversing the inequality sign
\be
K_1-K_2>0
\ee
Since case (c) arises for $K_1-K_2<0$, we see that to get this case we will need to  violate (\ref{acondition}) for some $w_i$. 

Note that if we have $w_i=-1$ for all $i$ then we get a Cosmological constant term in the Einstein equations, and thus we get exponential evolution. Though this is a faster expansion than the power laws we found for case (b), we still do not get a singularity at finite $t$.
 
\subsection{Absence of singularity at finite $t$ in case (a)}\label{aa4}

The physical part of the graph in Fig.\ref{parabolas}(a) ends at a finite value of $\tau$, but we would like to know if this corresponds to a finite or infinite value of the physical time parameter $t$. We will show that for physical values of parameters the point $\tau=\tau_2$ cannot be reached at finite $t$.

First we show this by a simple argument using the equation (\ref{qthree}), which gives
\be
\log \left ({\qth\over \qth_0}\right )=-\int_{t_0}^t {\sh\over \ih} dt
\label{logeq}
\ee
Recall that $\ih(t=t_0)>0$, and from (\ref{qfive}) and (\ref{ppos2}) we know that it cannot decrease. Thus the denominator of ${\sh\over \ih}$ is bounded below. On the other hand from (\ref{msix}) we see that $|\sh|$ is bounded for our interval $r_1\le \tau\le r_2$. Thus ${\sh\over \ih}$ is bounded for our evolution from $\tau=0$ to $\tau=r_2$.  But from (\ref{para}) we see that at the root $r_2$ we must have $\qth=0$ (since $\ih$ is bounded below and cannot vanish). So we see that the left hand side of (\ref{logeq}) must diverge for $\tau=r_2$, but from the right hand side of (\ref{logeq}) we find that this divergence cannot occur at any finite $t$.

Now let us establish this behavior directly by proving (\ref{alpharel}), which is
\be
{-{2(-r_2 K_2 + A_2)\over (K_1+K_2) ( r_1-r_2)}}+1<0
\label{tbe}
\ee
In the present case we have $K_1+K_2<0$, and from our ordering of roots we have $r_1-r_2<0$, so we can multiply the inequality (\ref{tbe}) by $(K_1+K_2)(r_1-r_2)$ without reversing the inequality sign
\be
-2(-r_2K_2+A_2)+(K_1+K_2)(r_1-r_2)<0
\ee
We write 
\be
\ph_0(\tau=r_1)\equiv \ph_1, ~~~\ph_0(\tau=r_2)\equiv \ph_2, ~~~\sh_0(\tau=r_1)\equiv \sh_1, ~~~\sh_0(\tau=r_2)\equiv \sh_2
\ee
From (\ref{mfive}),(\ref{msix}) we get
\be
-r_2K_2+A_2=\sh_2, ~~~~K_1(r_1-r_2)=\ph_1-\ph_2, ~~~K_2(r_1-r_2)=\sh_2-\sh_1
\label{fa3}
\ee
and the inequality to be established becomes
\be
(\ph_1-\ph_2)-(\sh_1+\sh_2)<0
\label{ineqs}
\ee

Consider the parabola $f(\tau)$ drawn in Fig.\ref{parabolas}(a). 
At the two roots $r_1, r_2$ the slope of $f(\tau)$ must be equal and opposite. Thus using (\ref{qsrel}) we have
\be
(\ph_1-\sh_1)=-(\ph_2-\sh_2)
\label{fa4}
\ee
Using this in (\ref{ineqs}) we find that the inequality to be established is
\be
-2\ph_2<0
\label{tbe2}
\ee
We have already seen in (\ref{ppos2}) that $\ph$ is nonnegative throughout the evolution. We will get $\ph(\tau_2)=0$ only if $\qth\sim\rho=0$ everywhere, which corresponds to empty Minkowski space and thus gives no singularity anywhere. For $\ph_2>0$ we get  
(\ref{tbe2}), which establishes (\ref{tbe}).

Using the notation above we can obtain (\ref{forapp}). From (\ref{forapp2}),(\ref{fa3}),(\ref{fa4}) we see that 
\be
{{2(\delta_k r_2+f_k)\over 2(-r_2 K_2 +A_2)-(K_1+K_2)(r_1-r_2)}}= {2\hat\gamma_k\over 2 \sh_2-(\ph_1-\ph_2+\sh_2-\sh_1)}={\hat \gamma_k\over  \ph_2}={\gamma_k\over \p_2}
\ee
which gives (\ref{forapp}).

\end{document}